# Space, Time and Machines

ARTO ANNILA[1,2,3,*]

[1]*Department of Physics,* [2]*Institute of Biotechnology and* [3]*Department of Biosciences, FI-00014 University of Helsinki, Finland*

**Abstract** The 2nd law of thermodynamics is used to shed light on present-day puzzles in cosmology. The universal law, given as an equation of motion, describes diverse systems when consuming free energy via various mechanisms to attain stationary states in their respective surroundings. Expansion of the Universe, galactic rotation and lensing as well as clustering of red-shifted spectral lines are found as natural consequences of the maximal energy dispersal that satisfies the conservation of energy, in the forms of kinetic, potential and dissipation. The Universe in its entirety is pictured as a giant Riemann resonator in evolution via step-by-step spontaneous breaking of one stationary-state symmetry to another to diminish energy density differences relative to its zero-density "surroundings". The continuum equation of evolution is proven equivalent to the Navier-Stokes equation. The ubiquitous flow equation has no solution because the forces and flows are inseparable when the dissipative process has three or more degrees of freedom. Since an evolving system is without a norm, there is no unitary transformation to solve the characteristic equation, but detailed trajectories remain inherently intractable. Conversely, stationary-state trajectories can be solved.

## 1. Introduction

Peculiarities promote physics. Presently it appears particularly puzzling that the Universe, as inferred from supernovae red-shifts [1,2,3,4], is expanding at an increasing rate – possibly due to dark energy. Also it is seen disconcerting that galaxies are rotating faster and bending light stronger than their detectable matter seems to account for [5,6,7] – possibly due to dark matter. Furthermore, it is intriguing that in the distant past the fine-structure constant might have differed from its present-day value based on measurements of the 21-cm hydrogen spectral line absorbed in molecular clouds along the line-of-sight to distant quasars [8,9,10]. Moreover, the Doppler shifted lines emitted from the far-away galaxies give the impression that red-shifts are clustered [11,12,13] – possibly due to some galactic-scale cooperative motions.

We are in no position to judge the accurateness of these peculiar findings or to question the correctness of the interesting explanations proposed for them, but here the aim is to acquire an overall understanding of the phenomena from the 2nd law of thermodynamics. In other words, the objective of this study is not to elaborate on critical assessments of the data or suggested interpretations but to provide a holistic view of nature in agreement with observations using the universal law.

Thermodynamics values everything in terms of energy hence also the ubiquitous natural law of least-time consumption of free energy can be expressed in general terms. It says: wherever there exists a difference of energy density, a flow of energy can appear to diminish that difference. The same meaning, with an incentive to utilize the energy difference, was already stated by Carnot: *wherever there exists a difference of temperature, motive force can be produced* [14]. A flow of energy is motion that naturally selects the fastest ways, i.e., the most voluminous steepest descents to level off the free energy landscape. In other words, when the system is evolving from one state to another toward the free energy minimum, entropy, as the additive probability measure of a system [15], will not only be increasing but it will be increasing in the least time.

The natural process of least-time free energy consumption, when written as an equation of motion derived from statistical physics of open systems [16], complies with conservation of energy: changes in kinetic energy equal changes in scalar and vector potentials (and vice versa). The basic relation between the three forms of energy was put forward already a long time ago [17,18]. Recently it has served to rationalize diverse inanimate, just as animate, evolutionary courses [19,20,21,22,23,24,25,26,27,28,29,30] and shed light on some central concepts, most notably space and time [31,32]. The irreversible energy transduction process, in the form of the 2nd law of thermodynamics, can also be formulated by the principle of least action and by Newton's 2nd law of motion [33,34].

In this study the old law will give nothing new as such, nonetheless, it will serve to put present-day peculiarities in perspective. When introducing basic concepts to address the perplexing phenomena, we will note the historical advent of ideas about evolution. Conclusions are corollaries of the 2$^{nd}$ law, the validity of which is itself not questioned here. No theory, by the words of Eddington, may go against it [35].

## 2. The energy dispersal process

The 2$^{nd}$ law of thermodynamics is communicated by general concepts, of which energy is essential. It can be assigned to anything that exists. Here we adopt Gibbs' notation [36] in assigning each entity $j$, present in indistinguishable numbers $N_j$, with an energy density $\phi_j = N_j\exp(G_j/k_BT)$ where $G_j$ is relative to the average energy per entity $k_BT$ of the system. According to the scale-independent formalism [37,38] each $j$-entity is a system of diverse $k$-entities. Each population $N_k$ is, in turn, associated with $\phi_k = N_k\exp(G_k/k_BT)$. Mutual interactions among these (material) entities (fermions) that exclude each other in space are communicated during time by force-carrier quanta (bosons).

We follow Boltzmann [39] in characterizing the pool of $j$-entities by the probability [16,31,32]

$$P_j = \left(\left(\prod_k N_k e^{-(\Delta G_{jk} - i\Delta Q_{jk})/k_BT}\right)^{g_{jk}} \Big/ g_{jk}!\right)^{N_j} \Big/ N_j! \quad (2.1)$$

defined in a recursive manner so that each $j$-entity is a product $\Pi N_k$ of embedded $k$-entities, each distinct type available in numbers $N_k$. The energy difference between the $j$ and $k$ entity is $\Delta G_{jk} = G_j - G_k$. The energy flux from the surroundings to the system is denoted by quanta $\Delta Q_{jk}$ that couple to the $jk$-transformation orthogonal to $\Delta G_{jk}$, hence indicated by $i$. For example, when momentum of an electron increases, light will emit perpendicular to the trajectory. In other words scalar and vector potentials are orthogonal to each other. The degeneracy $g_{jk}$ numbers the $k$-entity copies that remain indistinguishable (symmetric) in $j$.

The total probability for the entire system housing a diversity of $j$-systems is defined in a self-similar manner as the additive, logarithmic measure [16,31,32]

$$\ln P = \sum_j \ln P_j \approx \sum_j N_j \left(1 - \sum_k A_{jk}/k_BT\right) \quad (2.2)$$

where the free energy $A_{jk} = \Delta\mu_{jk} - i\Delta Q_{jk}$, i.e., affinity [40], is the motive force that directs the transforming flow $dN_j/dt$ from $N_k$ to $N_j$. The logarithmic density difference is given as the scalar potential difference $\Delta\mu_{jk} = \mu_j - \Sigma\mu_k = k_BT(\ln\phi_j - \Sigma g_{jk}\ln\phi_k/g_{jk}!)$. Stirling's approximation $\ln N_j! \approx N_j\ln N_j - N_j$ implies that the $j$-system is able to absorb or emit quanta without a marked change in its average energy content $A_{jk}/k_BT \ll 1$. Otherwise, e.g., when the population $N_j$ goes extinct or emerges, $\ln P_j$ is not a sufficient statistic for $k_BT$ [41]. An ensemble of entities where mutual interactions are insufficient to establish a common $k_BT$, is not defined as a system but regarded as surroundings of sufficiently statistic systems at a lower level of hierarchy [42,43] (Fig. 1).

The free-energy measure (Eq. 2.2) relates to entropy by the definition $S = k_B\ln P$. A system that is higher in energy density than its surroundings will evolve from its current state to another more probable one by displacing quanta to its surroundings. Likewise, a system that is lower in energy density than its respective surroundings will evolve by acquiring quanta from the surroundings. The states are distinguishable from each other only when the $jk$-transformation is dissipative $\Delta Q_{jk} \neq 0$ [16,31,33]. Systems evolve toward stationary states where their energy densities equal that of their surroundings. A system without net fluxes to or from its surroundings is in its entirety at a stationary state ($d\ln P = 0$). Its motions are reversible, conserved currents on tractable trajectories [44]. Likewise, a closed system that is cut off from its surroundings does not face evolutionary forces and does not evolve either.

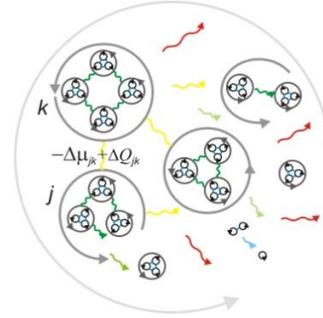

Figure 1. Self-similar diagram depicts energy transduction in a nested hierarchy of interacting systems within systems where each system (loop) is immersed in its surrounding system (larger loop). The mutual energy density differences and relative to the surroundings contained in $\Delta\mu_{jk}$ and $\Delta Q_{jk}$ (color-coded wavy arrows) drive the $j$ and $k$ systems (open loops) in evolution toward more probable states.

The equation of motion [16] for the open, evolving system is obtained from Eqs. 2.1 and 2.2 by differentiating $(\partial P_j/\partial N_j)(dN_j/dt)$

$$\frac{dS}{dt} = k_B\frac{1}{P}\frac{dP}{dt} = k_B L \geq 0 \quad (2.3)$$

where the generator of the natural process is

$$L = -\sum_{j,k} \frac{dN_j}{dt} \frac{A_{jk}}{k_B T}. \quad (2.4)$$

The time step $dt$ is denoted as continuous but the actual transformations advance in discrete steps. Moreover, when the quantized process depends on path, the time differential is properly denoted as inexact $đ$.

According to the principle of maximal entropy production [45] and the principle of least time [46] and the maximum power principle [47] the system will consume free energy as soon as possible. The energy flows direct from high $k$- to low $j$-densities along the steepest gradient $\partial/\partial t$ in time. This natural selection for the maximal dispersal manifests itself as the most voluminous flows along the steepest directional descents $v_j \partial/\partial x_j$ in space. The equivalence between the temporal and spatial directional derivatives follows from the conservation of energy. The balance between the scalar and vector potentials is contained, e.g., in gauge invariance [31].

Importantly, flows and forces in the process generator $L$ (Eq. 2.4) cannot be separated from each other when there are alternative $jk$-paths, i.e., three or more degrees of freedom. Therefore the equation of motion (Eq. 2.3) cannot be integrated to a closed form to predict non-holonomic evolutionary trajectories. During the self-similar processes populations will grow in sigmoid manner, i.e., mostly following the power law [24]. When the transduction mechanisms remain invariant, the flow is proportional

$$dN_j/dt = -\sum_{j,k} \sigma_{jk} A_{jk}/k_B T \quad (2.5)$$

to the free energy, i.e., affinity $A_{jk}$ by conductance $\sigma_{jk}$. However, due to the energy intake from surroundings new means $\sigma_{jk}$ may emerge and old ones may go extinct.

A differential change in $P$, also referred to as the coupling parameter $g$ [48], due to a change in the energy scale $k_B T$ is obtained from Eq. 2.2 as $C = TdS/dT = k_B \partial \ln P / \partial \ln T$. This formula for the specific heat is also known as the renormalization group equation, or simply the $\beta(g)$-function [49]. A stationary system has $C > 0$, as usual, whereas when $\Delta Q$ from surroundings exceed $\Delta \mu$ the evolving system will show $C < 0$. For example, a star that has exhausted its nuclear fuel will contract and heat up [50]. In the gauge theory these characteristics of $C$ at high energies are related to the asymptotic freedom [51].

The physical portrayal of natural selection for the maximal energy dispersal is no different in biological contexts but it is communicated differently [52]. Diverse energy transduction mechanisms are referred to as species. The most effective ones in consuming free energy are referred to as the fittest [23]. Biosphere with its diversity of mechanisms is the machinery for energy dispersal that generates numerous flows from the energy difference between the hot sun and cold space. The global system evolves in the quest for a steady-state partition of populations which is the most effective in transduction [25].

## 3. On the expansion of the Universe

According to the scale-independent 2$^{nd}$ law the Universe, as any other system, is evolving by dispersing energy from high to low densities in the least time to attain a stationary state in its zero-density "surroundings". Diverse celestial mechanisms, stars, black holes, quasars, galaxies, *etc*., make the machinery for irrevocable dilution of energy densities. The flow of energy from high to low spatial densities has been identified with the flow of time [31]. In the quest to disperse in least time, the high-density transformers displace from each other, on the average, as far away as possible. Thus the cosmological principle is seen as a compulsory consequence of the natural selection for the maximal dispersal [30]. *Nota bene*, the objective of this study is not to elaborate *how* particular mechanisms of the natural processes actually accomplish the maximal dispersal but only to address the principle in question as to *why* the Universe is expanding.

The nearly perfect thermal black-body spectrum of the cosmic microwave background radiation implies high homogeneity at the largest scale [53]. This furthers one to inspect the 2$^{nd}$ law (Eq. 2.3) defined for the continuum

$$T\frac{dS}{dt} = k_B T \frac{d(\ln P)}{dt} = -\sum_{j,k} \frac{dx_j}{dt}\frac{\partial U_{jk}}{\partial x_j} + i\sum_{j,k}\frac{\partial Q_{jk}}{\partial t} \quad (3.1)$$
$$\Rightarrow \sum_{j,k} \frac{d}{dt} 2k_{jk} = -\sum_{j,k}\frac{dx_j}{dt}\frac{\partial u_{jk}}{\partial x_j} + i\sum_{j,k}\frac{\partial q_{jk}}{\partial t}$$

where the free energy ($A_{jk}$) density $A_{jk}$ is composed of the potential energy ($U_{jk}$) density $u_{jk}$ and the dissipation ($Q_{jk}$) in the radiation density $q_{jk}$ along the temporal coordinate $t$ orthogonal to the spatial coordinates $x_j$ that are customarily given by combinations of Cartesian components $j,k = \{x,y,z\}$. The flow equation says: when scalar potential density $du_{jk}$ is consumed during $dt$, dissipation $dq_{jk}$ is produced and the balance is maintained by a change in the kinetic energy density $d(2k_{jk})$ (Fig. 2).

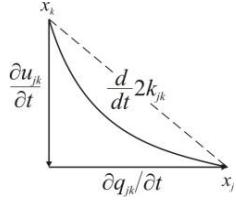

Figure 2. Flow of energy directs from the high potential density $u_k$ at $x_k$ down along the geodesic toward the low density $u_j$ at $x_j$ displaced by concurrent dissipation $\partial q_{jk}/\partial t$. The arc can be approximated by the chord (dashed line) corresponding to Lorentzian metric in a locality where the non-Euclidean energy landscape does not curve much.

The balance in the flows of energy (Eq. 3.1) is also available from the Newton's 2$^{nd}$ law of motion that defines the force density $\mathbf{f} = d\boldsymbol{\pi}/dt = d(\rho\mathbf{v})/dt$ by the change in the momentum density $\boldsymbol{\pi}$. When $\mathbf{f}$ is differentiated respect to both mass density $\rho$ and velocity $\mathbf{v}$, and the product with $\mathbf{v}$ is taken, the conservation in the component form follows as

$$\sum_{j,k} v_j f_{jk} = \sum_{j,k} v_j \rho_{jk} a_k + i\sum_{j,k} v_j \frac{\partial \rho_{jk}}{\partial t} v_k \quad (3.2)$$
$$\Leftrightarrow \sum_{j,k} \frac{d}{dt} 2k_{jk} = -\sum_{j,k} v_j \frac{\partial u_{jk}}{\partial x_j} + i\sum_{j,k} \frac{\partial q_{jk}}{\partial t}.$$

The change in the kinetic energy density $2k_{jk} = v_j \rho_{jk} v_k$ equals the energy density changes in potential and dissipation. The irrotational gradient $\partial u_{jk}/\partial x_j = -\rho_{jk} a_k$ is defined as usual for $j,k = \{x,y,z\}$. The divergence-free part of the field, i.e., the gradient of the vector potential yields the dissipated density $\partial q_{jk}/\partial t = v_k(\partial \rho_{jk}/\partial t)v_j$. It stems from the changes in the mass density. The change in the metric tensor of the energy landscape is customarily given by the energy equivalent $d\rho_{jk}c^2 = dE_{jk} = n^2 dq_{jk}$ that can be radiated in the respective surroundings defined by the isotropic index of refraction $n = c/v$. The mass change is often ignored but it signifies the changes in interactions when the system evolves from one state to another. Thus, in this study the mass is understood by $E = mc^2$ only as a convenient way to denote a stationary system's the energy content that can eventually be radiated entirely to the surrounding vacuum. The dispersal of energy, hence also the propagation of information [54] and cause [31], is limited to the speed of light $c$. The emitted energy is usually denoted as a spatial gradient of Poynting's vector [55]. The dissipation directs along the vector potential that is orthogonal to the scalar potential.

According to the 2$^{nd}$ law (Eq. 3.1) the flow from the density at $x_k$ toward the density at $x_j$ couples inevitably with dissipation $q_{jk}$ when $x_k$ and $x_j$ are distinguishable from each other. In geometric terms the flow during $t$ from $x_j$ to $x_k$ along a directional arc $s_{jk} = s_{kj}^*$ makes an affine (*cf.* affinity), i.e., a causal connection between the two spatial density loci (Fig. 2). The arc length along a continuous curve $x = x(t)$ is $s = (\mathbf{s}^*\mathbf{s})^{½} = \int(\mathbf{F}\cdot\mathbf{v})^{½}dt = \int(d2K/dt)^{½}dt$ where the integrand is referred to as the Rayleigh-Onsager dissipation function [56] or $TdS/dt$ by Gouy and Stodola [57,58].

According to the principle of least action the arc on the energy density landscape is the geodesic for the maximal dispersal [31]. The non-Euclidean landscape is evolving by flows of energy that are changing with changing free energy. Consequently, when there are three or more degrees of freedom, flows will be intractable. However, when the gradient reduction over $dx$ during $dt$ is taken as infinitesimal, the path $ds$ will be a straight space-time line $ds = dx - icdt$. By this Euclidean approximation due to Gauss, the local landscape is a plane [59] so that the Pythagoras theorem holds $ds^2 = d\sigma^2 + c^2dt^2$. The conservation of energy and mass-energy equivalence identifies the proper length $d\sigma$ [60,61] to the infinitesimal change $du$ and $cdt$ to the dissipation $dq$ in the radiative transition. Thus the Lorentzian manifold $d\sigma^2 = ds^2 - c^2dt^2$ with its opposite signs for the spatial and temporal intervals follows from the continuity in the flows of energy (Fig. 2). Accordingly, the proper time $d\tau = id\sigma/c$ is zero for light propagating at a constant frequency along an isergonic path defined by the Lorenz gauge [62]. Conversely frequency $f$ is shifting during passage along a curved geodesic in order to balance changes in the surrounding density with index $n$ of refraction relative to the universal reference density [29].

In the limit of Euclidean continuum an infinitesimal change in $d(2k) \to ds^2$ balances $du \to c^2d\tau^2$ and $dq \to c^2dt^2$ and yields the famous invariant of motion [63]

$$d2k/dt = -\partial u/\partial t + \partial q/\partial t \Rightarrow ds^2 = -c^2d\tau^2 + c^2dt^2 \quad (3.3)$$
$$\Rightarrow d\tau/dt = \gamma = \sqrt{1 - v^2/c^2}.$$

Thus the Lorentzian metric and the factor $\gamma$ of the special relativity are found from the 2$^{nd}$ law as a consequence of the energy balance where the hypotenuse $d(2k)$ is related with the two sides $du$ and $dq$. In contrast Galilean transformation does not comply with the conservation of energy and the Abelian group of symmetry does not account for natural processes that are directional in decreasing the free energy by symmetry breaking. The dissipative transformations obey non-commutative algebra when breaking one stationary-state symmetry for another [64].

The conservation of energy (Eq. 3.3) during radiative transformations requires that the diminishing sources of radiation, i.e., the diverse dissipative mechanisms are displacing from each other with increasing average velocity $v$ approaching the speed of light ($\beta^2 = v^2/c^2 \to 1$). According to the integral form of Eq. 3.3 the Universe at the age of $t$ is expanding at the rate of the Hubble parameter $H = 1/t$ [65,66]. When the matter density $\rho$ is decreasing (Fig. 3), the change rate of expansion $dH/dt = -H^2$ is no peculiarity but a natural consequence of the conservation of energy among potential, dissipative and kinetic components. The least-time dispersal of energy shows up, for example, so that apparent magnitudes of standard candles fall faster than their spectra shift to red [29].

The universal energy density landscape that sums up from the spatial and temporal gradients is flattening toward the zero density of nothing. During the process the kinetic energy density will comprise an ever-larger fraction of the total energy of the ever-diluting Universe. Each evolutionary epoch will end when its characteristic machinery of energy dispersal has exhausted the free energy in its access and faced extinction. Since light that is characterized by the most elementary symmetry group cannot be broken down any further, it is the boundary of the material density manifold. Eventually, when approaching the zero-density equilibrium, all anisotropies in material forms will evaporate to mere radiation of vanishing density [67].

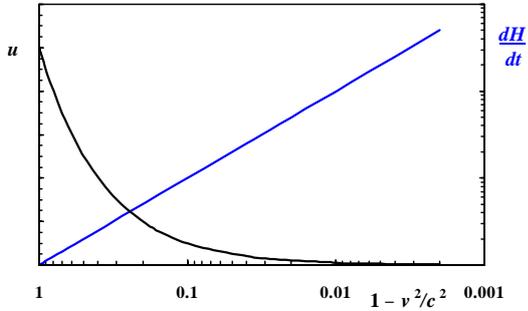

Figure 3. Combustion of bound energy density $u$ to free radiation, i.e., bosonic density $q$ results in the universal expansion where the balance between $u$ and $q$ is maintained when the devoured material sources are dispersing with increasing kinetic energy so that the average velocity $v$ is approaching the speed of light $c$. This manifests itself as a changing rate of expansion $dH/dt$.

The universal law of the maximal energy dispersal may at first sight appear as if it would give no role to gravitation. However, the gravitational potential is in Eq. 3.1 by the equivalence principle as $u_{jk} = \int v_j(\partial u_{jk}/\partial x_j)dt = -\int v_j\rho_{jk}a_k dt$. The gravitational potential is the density difference between $m_j$ at $x_j$ and $m_k$ at $x_k$ relative to the universal surroundings [34].

The familiar inverse distance dependence in $U$ follows from the stationary action (see appendix). During the diluting expansion the total gravitational potential is diminishing at the rate $-\partial U/\partial t$ because the total matter density is transforming to the dispersing dissipation $\partial Q/\partial t$. Thus at any time the total gravitational potential $U = GM^2/R$ equals energy $Mc^2$ in the total matter $M$ that is spun over the universal radius $R$ [68,69,70,71].

The nascent ($\beta = 0$) Universe is characterized by the acceleration $a_P = c/t_P = c^2/R_P$ in natural units [72] of time $t_P = 1/H_P = \sqrt{G\hbar/c^5}$ and radius $R_P = \sqrt{G\hbar/c^3}$ where $G$ is the gravitational constant and $\hbar$ is the Planck constant. The early, compact Universe is expanding at the rate $H_P = E_P/\hbar$ and fading with power $L_P = \hbar H_P^2 = -\hbar dH_P/dt$ toward the perfect flatness ($\beta \to 1$) and complete darkness ($f \to 0$) where $a_\infty = c^2/R_\infty = c/t_\infty = 0$. During the universal process $H = \dot{a}(t)/a(t)$ is changing as the scale factor [59] $a(t) = \sqrt{c^2\varepsilon_o\mu_o}$ is stretching from the reference index of refraction $n = a(t)^{-1} = 1$ defined by the present-time $t$ permittivity $\varepsilon_o$ and permeability $\mu_o$. Today the radius of space is huge $R = ct$. Its curvature corresponds to a tiny but non-negligible acceleration $a_t = GM/R^2 = 1/\varepsilon_o\mu_o R = c^2/R = c/t = cH$ due to the total mass $M$ that is diluted to the average density $\rho_t = 1/2\pi Gt^2$, estimated $9.9\cdot10^{-27}$ kg/m$^3$ [73].

## 4. On the rotation of galaxies

A galaxy with its stars, black holes, *etc*., is according to the 2$^{nd}$ law, in its entirety machinery to disperse energy, e.g., by nuclear reactions. At all times and levels of the nested organization of systems within systems [74] the kinetic energy balances the potential energy and dissipation in diverse transformations from high to low densities (Eq. 3.1). A detailed galactic kinematics of translational, orbital and rotational motions is an intricate many-body problem beyond the objective of this study. Here, the galaxy rotation problem is addressed only to understand the principle reason for the puzzling discrepancy between the observed orbital velocities of luminous stars or gaseous matter in spiral galaxy disks and the estimates provided by the gravitational potential considering only the visible mass.

The energy balance by the 2$^{nd}$ law (Eq. 3.1)

$$d2K/dt = -v\,\partial U/\partial r + \partial Q/\partial t$$
$$\Rightarrow 2K = -U + Q \Rightarrow v^2 = GM_O/r + Q/m \qquad (4.1)$$

yields the steady-state velocity $v$ formula for a body of mass $m$, subject to the gravitational field $a = GM_O/r^2$ at the orbit $r$ that encloses a central mass $M_O$. For non-dissipative ($Q \approx 0$) structures, such as planets, integration of Eq. 4.1 reproduces the familiar stationary-state virial theorem $2K + U = 0$.

Very far away from the galactic center, at the spatial coordinate $r_t$ well beyond the luminous edge, the galactic gravitational tug $a = GM_o/r_t^2$ is feeble. There the universal acceleration $a_t = GM/R^2 = c^2/R = v^2/r_t$ curves the energy landscape according to Kepler's third law (Fig. 4). When $a_t/v^2 = 1/r_t$ is used in Eq. 4.1 $mv^2 = GmM_o/r_t = a_tGmM_o/v^2$ a gas cloud of mass $m$ is found to orbit according to the Tully-Fisher relation $v^4 = a_tGM_o$ that is the asymptote of galactic rotational curves [75]. Earlier, the $v^4$-form has been worked out from the modified Newtonian dynamics [76]. Here the tiny acceleration $a_t = cH_o/2\pi$ proportional to the Hubble constant $H_o$ is no anomaly [77,78,79] or attributed to radiative momentum transfer [80] but identified by the 2$^{nd}$ law to the curvature caused by the non-zero universal energy density.

The above derivation of the $v^4$-form though, may appear odd for some because the *universal* acceleration contributes to the rotation about a *local*, galactic center. However from the thermodynamic viewpoint, the Universe with galaxies is no different from other cases where a global gradient, such as an overall temperature gradient between a warm ocean and the cold atmosphere, or across a Benárd cell or an overall velocity gradient in a Taylor-Couette cell, generate local convection whirls [81,82] or where a total centrifugal acceleration in a superfluid or a total magnetic field in a superconductor produce local quantized vortices [83].

Consistently with the conservation of energy, the integral form of the 2$^{nd}$ law, i.e., the principle of least action [84,33] requires that the action $Qt$ in dissipation will open the path $\int (d2K/dt)^{1/2} dt$, e.g., a perihelion [85] will advance, from that closed, modular orbit determined by the conserved action $Ut$ of the scalar potential alone [30]. Likewise, the energy conservation requires that a meniscus of water, i.e., the potential energy surface curves to balance the kinetic energy due to the rotation of a bucket. The universal curvature contributes in many motions but today being so small it is observable only at large scales, e.g., involving galactic clusters [86,87]. In the distant past the stronger universal force played a more significant role, e.g., in laboring anisotropies. Their skewed distributions [88] that accumulate in a power-law manner signal that the dispersal progressed according to the 2$^{nd}$ law.

The energy balance (Eq. 4.1) for dissipative structures ($dQ \neq 0$), such as stars, black holes and galaxies, where the kinetic energy tallies with potential energy and dissipation results in motions that manifest not only as orbital but as translational motions, axial spinning and convections. Despite complex dynamics, a fraction of $2K$ due to the dissipation $Q$ is expected to contribute to the orbital velocities $v$ beyond that due to the scalar potential $U$ alone. For example, the luminosity of Sun $\partial Q/\partial t$ is not negligible at its orbit $r$ about the Milky way's center but comparable to the galactic attraction $-v\partial U/\partial r$. Differences in velocities due to differences in dissipation give rise to dynamics. It is most apparent between dissipative and non-dissipative structures, e.g., when a newborn star lights up, it will due to emitted power $\partial Q/\partial t$ leave its nursery, Bok globule [89]. The velocity differences between stars of differing luminosities are not that dramatic. When $dm$ is combusted during $dt$, Eq. 4.1 yields $dQ/m = v^2 dm/m = v^2 d\ln m$ in accordance with the mass-luminosity relationship that is approximately a power-law for the main-sequence stars. All in all it is concluded that the galactic rotation curve is not a peculiarity but a natural consequence of the energy conservation in dispersal.

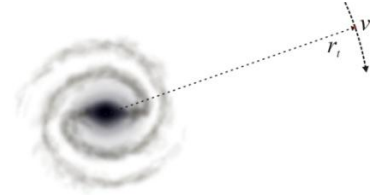

Figure 4. Far away from the center of a galaxy with mass $M_o$ the energy density landscape experienced by a body $m$ is dominated by the universal curvature $1/R = a_t/c^2$ so that $c^2/R = v^2/r_t$ and the balance $mv^2 = GmM_o/r_t$ between the kinetic and potential energy enforces the orbital velocity $v^4 = a_tGM_o$.

## 5. On the flow equation

The reasoning based on the 2$^{nd}$ law of thermodynamics about the diverse celestial motions may seem as if intricate dynamics were overlooked. However, the equation of motion (Eq. 2.3) is in its simplicity an insightful account on complexity. To begin with, the flow equation cannot be solved analytically when the driving forces $A_{jk}$ and flows $v_j$ are inseparable [16]. This is the case when there are three or more degrees of freedom for the dissipative processes [90]. Intractable, even chaotic tracks are familiar from the three-body problem [91] because in three or more dimensions Poincaré-Bendixson theorem does not apply [92]. Non-deterministic motions are the flows of energy with degrees of freedom between the system and its surroundings that level the density differences. The dissipation term in Eqs. 2.1, 2.3, 3.1 and 4.1 captures the role of surroundings to the evolution of a system. Non-determinism was phrased by Eddington: *It is impossible to trap modern physics into* *predicting anything with perfect determinism because it deals with probabilities from the outset* [93].

To gain further insight to the myriad of motions, the 2$^{nd}$ law as the continuum flow equation (Eq. 3.1) is related to Navier-Stokes equation. The Navier–Stokes equation states the conservation of momentum density $\Sigma_{jk}\pi_{jk} = \Sigma_{jk}\rho_{jk}v_k$ in flows along directional spatial $v_j\partial/\partial x_j$ and temporal gradients $\partial/\partial t$ for the Cartesian combinations $j,k = \{x,y,z\}$.

**Theorem 5.1** *The Navier-Stokes equation is Newton's 2$^{nd}$ law of motion.*

*Proof.* Newton's 2$^{nd}$ law of motion, defined as $\mathbf{f} = d(\rho\mathbf{v})/dt$ for the force $\mathbf{f}$ and momentum $\boldsymbol{\pi} = \rho\mathbf{v}$ density, is in the component form $j,k = \{x,y,z\}$

$$\sum_{j,k} f_{jk} = \sum_{j,k} \frac{\partial}{\partial t} \rho_{jk} v_k . \qquad (5.1.1)$$

To keep the notation simple, the law of motion (Eq. 5.1.1) is redundantly multiplied by 2 so that a temporal gradient $\partial/\partial t$ can be rewritten as an equivalent directional spatial gradient $v_j\partial/\partial x_j$ using the conservation of energy to obtain

$$2\sum_{j,k} f_{jk} = \sum_{j,k}\left(\frac{\partial \rho_{jk}}{\partial t}v_k + \rho_{jk}\frac{\partial v_k}{\partial t}\right) + \sum_{j,k}\left(v_j\frac{\partial \rho_{jk}}{\partial x_j}v_k + v_j\rho_{jk}\frac{\partial v_k}{\partial x_j}\right). (5.1.2)$$

The conserved and non-conserved ($\partial\rho_{jk}/\partial t \neq 0$) terms in 5.1.1 are regrouped

$$2\sum_{j,k} f_{jk} = \sum_{j,k}\left(\rho_{jk}\frac{\partial v_k}{\partial t} + v_j\rho_{jk}\frac{\partial v_k}{\partial x_j}\right) + \sum_{j,k}\left(\frac{\partial \rho_{jk}}{\partial t}v_k + v_j\frac{\partial \rho_{jk}}{\partial x_j}v_k\right). (5.1.3)$$

The non-conserved terms are rearranged and combined to

$$2\sum_{j,k} f_{jk} + \sum_{j,k}v_j\left(\rho_{jk}\frac{\partial v_k}{\partial x_j}\right) = \qquad (5.1.4)$$
$$\sum_{j,k}\rho_{jk}\left(\frac{\partial v_k}{\partial t} + v_j\frac{\partial v_k}{\partial x_k}\right) + \sum_{j,k}\left(\frac{\partial \rho_{jk}}{\partial t}v_k + v_j\frac{\partial}{\partial x_j}\rho_{jk}v_k\right).$$

Finally, the component form of Eq. 5.1.4 is abridged to the vector notation

$$\mathbf{b} = \rho\left(\frac{\partial \mathbf{v}}{\partial t} + \mathbf{v}\cdot\nabla\mathbf{v}\right) + \mathbf{v}\left(\frac{\partial \rho}{\partial t} + \nabla\cdot(\rho\mathbf{v})\right) \qquad (5.1.5)$$

which is the Navier-Stokes equation in its general form, known also as the Cauchy momentum equation. The generic body force density $\mathbf{b} = \Sigma_{jk}(v_j\rho_{jk}\partial v_k/\partial x_j + 2f_{jk})$ is referred to as a source (or sink) of momentum density where the first term is recognized as the stress tensor $\boldsymbol{\sigma}$ when denoting $\partial v_k/\partial x_j = a_k$ and writing the potential energy gradient per area $A$ as $\Sigma_{jk}\rho_{jk}a_k/A = -\Sigma_{jk}\partial u_{jk}/\partial x_j/A = \nabla\cdot\boldsymbol{\sigma}$. The trace of diagonal elements (normal stresses) $\frac{1}{3}\Sigma\sigma_{jj} = -p$ equates with pressure and the off-diagonal elements ($j\neq k$) (shear stresses) embody the traceless deviatoric stress tensor $\mathbb{T}$. Often, but not in this study, the non-conserved term on the right hand side is assumed to be zero by the mass continuity $d\rho = 0$. Such a simplification neglects changes in interactions due to dissipation and is thus motivated only in studies of conserved (stationary) systems.

**Theorem 5.2** *The Navier-Stokes equation is the 2$^{nd}$ law of thermodynamics in the continuum.*

*Proof.* When the Navier-Stokes equation, as in Eq. 5.1.3, is multiplied by $v_j$ and the identity $\partial/\partial t = v_j\partial/\partial x_j$ is used

$$2\sum_{j,k}v_j f_{jk} = \sum_{j,k}v_j\left(\rho_{jk}\frac{\partial v_k}{\partial t} + v_j\rho_{jk}\frac{\partial v_k}{\partial x_j}\right)$$
$$+ \sum_{j,k}v_j\left(\frac{\partial \rho_{jk}}{\partial t}v_k + v_j\frac{\partial \rho_{jk}}{\partial x_j}v_k\right) \qquad (5.2)$$
$$\Leftrightarrow 2\sum_{j,k}v_j f_{jk} = 2\sum_{j,k}v_j\rho_{jk}\frac{\partial v_k}{\partial t} + 2\sum_{j,k}v_j\frac{\partial \rho_{jk}}{\partial t}v_k$$
$$\Leftrightarrow \sum_{j,k}\frac{\partial}{\partial t}2k_{jk} = -\sum_{j,k}v_j\frac{\partial u_k}{\partial x_j} + \sum_{j,k}\frac{\partial q_{jk}}{\partial t} ,$$

the 2$^{nd}$ law of thermodynamics is obtained as defined by Eq. 3.1 using the definitions of the kinetic $\Sigma v_j f_{jk} = \Sigma v_j\partial(\rho_{jk}v_k)/\partial t = -\Sigma\partial(2k_{jk})/\partial t$ and potential energy density $\rho_{jk}\partial v_k/\partial t = -\partial u_{jk}/\partial x_j$ and the density in dissipation $\partial q_{jk}/\partial t = v_j(\partial\rho_{jk}/\partial t)v_k$.

**Theorem 5.3** *In three space dimensions and time, given an initial non-zero velocity field, there exists no analytical solution for a non-zero vector velocity and a scalar pressure field that would solve the Navier-Stokes equation.*

*Proof.* According to the theorem 5.2 the Navier-Stokes equation is the continuum flow equation known as the 2$^{nd}$ law of thermodynamics. It follows from the properties of the 2$^{nd}$ law that the equivalent Navier-Stokes equation has no analytical solution, when there are three or more degrees of freedom, for a given initial velocity field by integration to a closed form for the vector velocity $\mathbf{v}(x,t)$ and a scalar pressure $p(x,t)$ field because the vector velocity components $v_j$, that appear as the kinetic energy changes $\Sigma_{jk}d(2k_{jk})/dt$, are inseparable in Eq. 3.1 from the force components, i.e., the pressure due to the scalar potential energy gradient $\Sigma_{jk}\partial u_{jk}/\partial x_j$ per area $A$ and vector dissipation $\Sigma_{jk}\partial q_{jk}/\partial t$. The dissipating system has no deterministic analytical trajectory because closed orbits are impossible according to the theorem for gradient systems with three or more dimensions [92]. When $\Sigma_{jk}v_j\partial u_{jk}/\partial x_j - \Sigma_{jk}\partial q_{jk}/\partial t \neq 0$, the trajectories are unbound because the kinetic energy density

$$\int_{t_0}^{t} \sum_{j,k} v_j \left( -\frac{\partial u_{jk}}{\partial x_j} + \frac{\partial q_{jk}}{\partial x_j} \right) dt = \int_{t_0}^{t} \sum_{j,k} v_j \frac{\partial}{\partial t} \rho_{jk} v_k dt \quad (5.3)$$

$$= \int_{t_0}^{t} \sum_{j,k} v_j \frac{\partial \pi_{jk}}{\partial t} dt = \int_{t_0}^{t} \sum_{j,k} \frac{\partial}{\partial t} 2k_{jk} dt > 0$$

is unbound unless $\Sigma_{jk}\rho_{jk} = 0$ or $v_j = 0$ for all $j,k = \{x,y,z\}$. Thus there is no constant $\varepsilon \in (0, \infty)$ for the gradient system with three or more degrees of freedom so that $\int v^2(x,t) dt < \varepsilon$ within $\mathbb{R}^3$ or equivalently so that $\int_{\mathbb{R}^3} |\mathbf{v}(x,t)| dx < \varepsilon$ for all $t$.

The lack of a closed form integral solution of the Navier-Stokes equation for three or more dimensions [94] is a direct consequence of the principle of increasing entropy $dS/dt > 0$. The energy $TdS/dt = k_B T d(\ln P)/dt = \Sigma d2K_{jk}/dt > 0$ of a system in evolution from one state to another by breaking symmetry is changing and the Noether's theorem of conserved currents [44] does not hold. Accordingly the vector velocity is unbound. The action $\int 2k dt = \int \rho v^2(x,t) dt$ cannot be integrated to a closed form either because energy is flowing from the system to the surroundings (or vice versa) until the stationary state has been attained. Since the non-conserved flow system has no norm, it cannot be represented by a unitary matrix (see Sec. 7). Consequently the eigenvalue problem cannot be solved to obtain the flow-field vectors. Thus, it is concluded the Navier-Stokes equation with three or more degrees of freedom does not have an analytical solution for a non-zero vector velocity and a scalar pressure field.

Conversely, when the number of degrees of freedom is two, there are no alternative paths. Hence the flow down from the source to the sink is tractable, although the force and the flow are interdependent, the problem is 1-separable and deterministic [90].

To gain further insight to the complicated motions, the master equation of evolution is rearranged from Eq. 2.3

$$\frac{dP}{dt} = LP \geq 0 \,. \quad (5.4)$$

When the surrounding energy density changes, driving forces will develop and the system will evolve (adapt) by diminishing the gradients. At critical transitions where changes in interactions, i.e., in mass ($d\rho_{jk} \neq 0$) are substantial, the conducting mechanism $\sigma_{jk} = \rho_{jk}^{-1}$ (Eq. 2.5) changes so that $k_B T$ is not a sufficient statistic for $\ln P_j$. When the non-conserved system emerges with a new (or looses an old) transduction mechanism, a path to intricate dynamics will open, and an exponent due to Kowalewskaja, associated with the new parameter, will appear for the first time [95]. The novel mechanism (e.g., tornado, turbulence, spark, species) is often regarded as surprising as its emergence cannot be deduced from the system's contemporary constituents alone. Indeed, the influx from the surroundings is the vital but often ignored ingredient of emergence, the interaction energy [96]. This reminds us from the quote by Eddington [97]: *We used to think that if we knew one, we knew two, because one and one are two. We are finding that we must learn a great deal more about 'and'.*

When dissipation is small compared with potential energy density ($d\rho_{jk} \approx 0$), $\ln P_j$ is a sufficient statistic for $k_B T$ and the flow $v_j \approx -\int \Sigma \sigma_{jk}(\partial/\partial x_j)(u-q) dt$ is smooth almost everywhere. Although the equation of evolution (Eq. 5.4) has in general no solution, it can be formally integrated in $n$ steps from an initial state $P_0 > 0$ to $P_n = P_0 \exp(nL)$. However, there is no guarantee for the global smoothness because the open system may encounter new driving forces on its way from one state to another, more probable one. Even chaotic motions may appear when $L > 0$ in accordance with a positive maximal Lyapunov exponent. Likewise, a positive eigenvalue of Jacobian means an unstable orbit [92].

When dissipation vanishes ($d\rho = 0$), $L = 0$. Accordingly the maximal Lyapunov exponent is not positive. Local motional modes of the landscape are customarily characterized by trace, determinant and discriminant of Jacobian [92]. The maximum entropy state is stable against perturbations $\delta$ in internal variables according to Lyapunov: $S(\delta) \leq 0$ and $dS(\delta(t))/dt > 0$ [98]. The closed orbits of the conserved currents are tractable, i.e., computable. The constant conductance $\sigma_{jk} = \rho_{jk}^{-1}$ means an invariant transduction mechanism [56]. Then the 2$^{nd}$ law reduces to $\rho d(v^2)/dt = -v du/dx$. The variables $v_j$ and $u_k$ can be separated and an analytical solution for the flow $v_j = -\int \Sigma \sigma_{jk}(\partial u_{jk}/\partial x_j) dt$ can be obtained from the Newton's 2$^{nd}$ law for the conserved force $\mathbf{F} = m\mathbf{a}$ by integration to a closed form. When the system is at the steady state with its surroundings, the vector velocity $\mathbf{v}(x,t) \in [C^\infty(\mathbb{R}^3 \times [0, \infty))]^3$ and a scalar pressure field $p(x,t) \in C^\infty(\mathbb{R}^3 \times [0, \infty))$ are smooth and globally defined.

Finally, we remark that according to the principal state equation (Eq. 2.1) the system is evolving in steps, not continuously. Due to the quantization also the vector velocity and the scalar pressure field are discrete functions. Evolution in time increments is often described by some recurrence equation $x_{t+1} = f(x_t)$ such as the logistic equation [99]. Iterative maps are illustrative in localizing points of bifurcations and regimes of chaos [100]. However, the master equation (Eq. 5.4) for the evolving probability is non-deterministic unlike, e.g., the logistic curve or law-of-mass-action models [101]. Therefore evolution in dissipative increments (Eq. 2.3) is amenable for numerical simulation where $k_BT$ and potentials are updated at each step.

## 6. On the spectrum of expansion

According to the 2$^{nd}$ law (Eqs. 2.3 and 5.4), the probability $P$ is increasing in discrete steps as energy is dispersing in quanta. The step-by-step evolution from one stationary state to another, more probable one, can be formulated by quantum mechanics in the same way as the continuous evolution by continuum mechanics.

Energy density of a quantized system is customarily described by a wave function $\psi(x,t)$. The probability current $dP/dt = d/dt \int \psi^*\psi dx$ [53] at $x$, in the middle of the transition from the state at $x_k$ to that at $x_j$, is obtained from the time-dependent equation of motion for $\psi$ using the fundamental theorem of calculus [31]. The energy balance in the transition is embedded in the familiar commutation relation $[\hat{p},\hat{x}] = px - xp = -i\hbar$. This form corresponding to $2Kt + Ut = Qt$ of an incoherent system follows from the quantized least action $(\partial/\partial t)(\hat{p}\hat{x}) = 0$ using the expectation values $\langle\partial\hat{p}/\partial t\rangle = -\partial U/\partial x$ for the momentum $\hat{p} = -i\hbar\partial/\partial x$ and coordinate $\hat{x} = x$ and energy $\hat{Q} = -i\hbar\langle\partial/\partial t\rangle$ operators. Transformations with net dissipation are non-commutative but when to-and-fro flows cancel each other exactly over the integration period, the system remains Hamiltonian, e.g., given by Liouville-von Neumann equation. Therefore the solutions are available by a unitary transformation to a time-independent frame which is in fact the signature of a stationary state. Operator algebra is a convenient way to track the relative phases of coherence pathways but it does not, as such, bring forward new evolutionary phenomena. Thus we prefer to use the basic formalism (Eqs. 2.1 – 2.4) to learn about the proposed galactic red-shift clustering and eventual variation in the fine-structure constant.

The initial state of the Universe at the ultraviolet fixed point is pictured by Eq. 2.1 as an elemental system $N_1 = 1$ composed of indistinguishable basic entities in degenerate numbers $N_0 = g_{10}$. The energy difference $\Delta G_{10} = G_1 - G_0 = k_BT$ between the nascent Universe and its "surroundings" at zero density $G_1 = 0$ is huge. The initial probability (Eq. 2.1) $P = [N_0\exp(-\Delta G_{10}/k_BT)]^{g_{10}}/g_{10}!$ is the natural starting point for a series of transformations to increase $P$ in least time.

The Universe will evolve by lowering its degeneracy from $g_{10}$ by spontaneous symmetry breaking in phase transitions where the basic entities will transform (combine) to diverse and distinct constituents in numbers $N_{k>0}$.

Concomitantly $\Delta G_{10}$ will fracture to diverse scalar $\Delta G_{jk}$ and vector $\Delta Q_{jk}$ potential differences. In other words, the high factorials and energy differences of the nascent improbable state are conquered by divisions as soon as possible. Recursive decimation-in-time by two is the familiar example from the fast Fourier transformation [102]. However the general principle does not expose particular mechanisms of dispersal, but the flows will naturally select those to disperse energy in least time. Independent of the short-scale dynamics the universal process is bound asymptotically to $P_{max}$. At the ultimate equilibrium i.e., at the infrared fixed point all densities will vanish ($k_BT \to 0$).

According to the step-by-step evolution, the Universe is an expanding resonator (cavity) where energy is dispersing to lower and lower harmonics [103]. The expression of propagating waves as a sum of standing waves subject to boundary conditions is familiar from the derivation of Planck's law [104]. During early epochs when the universal index of refraction $n$ was high, photons were not effective in dispersal until recombination [105] since high-energy fluxes suffer more than low-energy fluxes from internal reflections at boundaries. Therefore, to inflate in least time, the natural selection for the maximal dispersal might have favored some primordial dissipative mechanisms other than photons. When the "surroundings" is at zero density, the condition of the maximal transmission in a stratified medium [106] is the geometric mean ($\sqrt{n}$) which implies a power-series reduction (Fig. 5). A series in powers of two has been proposed to account for the quantized red-shifts of 21-cm lines recorded from distant galaxies [103,107,108].

As the universal resonator grew larger, its discrete energy density spectrum began to resemble more and more the famous continuum limit $q(f) = (8\pi hf^3/c^3)/[\exp(hf/k_BT) - 1]$ obtained by Planck. Currently at $T = 2.725$ K the integrated radiation density $q = \pi^2 k_B^4 T^4/15\hbar^3 c^3$ is about $6 \cdot 10^{-5}$ of $u$ [73] and balanced by $2k$ of the sources that are, on the average, dispersing with high velocity from each other. The skewed spectrum $q(f) \to 0$ as an example of a natural distribution is shifting to lower frequencies $f$ and falling in amplitude to remain in balance with the diminishing $u$ as $R = ct \to \infty$. The dilution is directing to the flat stage of nothing ($a_\infty = 0$, $k_BT = 0$) where no finite frequency photon is propagating either.

According to the step-by-step evolution, the light from a far-away source is not only reporting to us from the past conditions but the clustering of Doppler shifted lines results from the incremental frequency shifting as the light travels through the space that is undergoing a series of expanding transitions. The light from the distant past accumulates its long wavelength in a series of steps on its way through the diluting space where each state is characterized by $n^2 = \varepsilon\mu/\varepsilon_o\mu_o$ relative to the present state. Thus it is concluded that the 21-cm hydrogen line absorbed in molecular clouds along the line-of-sight to distant quasars does not signal from a varying fine-structure constant $\alpha = e^2Z/2h$ but from the light's path through the energy density that has been diluting in steps where the squared impedance of free space $Z^2 = \mu/\varepsilon = \mu_o/\varepsilon_o$ remains a constant. Likewise, the clustering of red-shifted lines emitted from the galaxies far-away does not signify galactic-scale collective phenomena but reveals that the intervening space as the energy landscape is flattening via the step-by-step expansion.

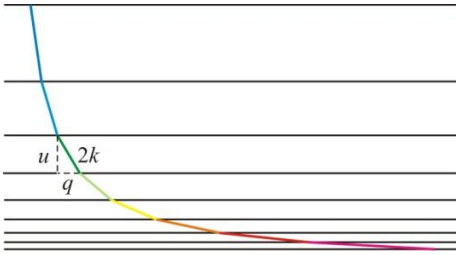

Figure 5. Evolving energy level diagram during the maximal dispersal of energy. At each dilution step $t \to t'$ the decreasing scalar potential density $u$ is transformed to photon density $q$ and balanced by the kinetic energy density $2k$ (color coded) so that the average energy density decreases by $\varepsilon_o\mu_o/\varepsilon_{t'}\mu_{t'}$ relative to the present permittivity $\varepsilon_o$ and permeability $\mu_o$.

The principle of least action reveals also that light balances changes in potential by changing its frequency. This is apparent when a ray is tracing past a local anisotropy [29]. For example, a beam of light when entering in a stratified medium, simplified to a continuum gravitational potential $U(r)$, will shift higher in frequency at each boundary and thus it will also bend more and more toward the field lines as it exposed to the higher and higher potential characterized by higher and higher index of refraction. Likewise, a beam of light, when exiting, will shift lower in frequency at each step and thus it will also bend more and more away from the field lines as it is subject to the lower and lower potential (see appendix). In other words, during the propagation of light, as in any other natural process, the driving force and the flow are interdependent. Thus the conservation of energy gives rise to the enhancing refraction through a centro-symmetric potential and drives the ray of light to a closer encounter at a stronger curvature and gives rise to a stronger lensing [109,110] than expected when the wavelength change, that is balancing the potential change, is neglected (Fig. 6). In general, the longest waves deflect least in a local and universal curvature and thus reach furthest.

Light that is spiraling in a gravitational field with no return is destined to a black hole. In relation to the associated information paradox, thermodynamics maintains that no information exists without physical representations. Therefore the thermodynamic entropy $S = k_B\ln P$ is the proper measure of information [54]. It follows that information is lost when its physical representations are swallowed and transformed by the black hole. It is a common tenet but incorrect thought that complete information about a physical system at one point in time would determine its state at any other time. Information, equated via $S$ with the free energy, does not suffice to determine future states of a dissipative system with degrees of freedom [90]. A path of a non-Hamiltonian system along diminishing free energy cannot be integrated in a closed form, i.e., predicted because the forces are inseparable from the flows. The surroundings have a say in the future course of any system. The black hole is an evolving system, as any other, diminishing energy density differences in its surroundings. The anisotropic distribution of surrounding supplies at a center of a galaxy makes the black hole a particularly effective mechanism for dispersal because it is absorbing from the crowed accretion disk and emitting in blazing transformation to the spacious axial void.

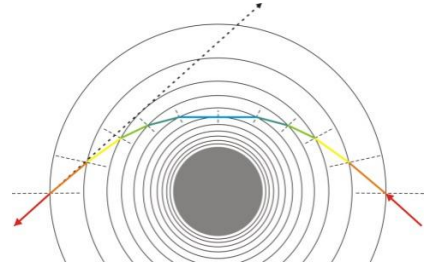

Figure 6. Drawing illustrates a ray of light through a centro-symmetric stratified potential (co-centric isergonic contours with the dashed field lines). The path traces a curved geodesic where frequency shifting (color coded) balances the changing scalar potential to satisfy the conservation of energy according to Fermat's principle. Thus the line-of-sight (dashed arrow) departs from the source direction.

## 7. On the Riemannian resonator

The description of an evolving system as a flattening energy landscape (Fig. 5) prompts one to represent states by phasors (phase vectors) whose amplitudes are proportional to the local curvature, i.e., inverse of the radius' length and whose directions are defined by the interdependent scalar $U$ and vector $Q$ potentials and kinetic energy $2K$. In general, phasors as a vector field span a curved affine space. In particular, the stationary-state phasors form the Hilbert space with (Euclidean) norm. This stationary set of inner product space can be represented by an algebraic number field so that each member is indexed by coprime integers $q$ and $k$, i.e., $\gcd(q,k)=1$ [111] (Fig. 7).

The landscape's radius of curvature will grow when one symmetry breaks spontaneously to another. The spectral radius will engulf a new unique stationary state represented by the phasor $z_k^{-q}$ that points along the direction $\exp(-i2\pi q/k)$ on the ring $\mathbb{Z}/k\mathbb{Z}$ of integers modulo $k$ with unit $q$ when the primitive root $z_k^{-qr} \equiv q \pmod{k}$ congruent to a power $r$ equals one. The discrete Fourier transformation of $\gcd(q,k)$ picks by the totient $\varphi(k)$ the number of elements in the group $\mathbb{Z}_k^\times$ of units $q$ with multiplication modulo $k$ [112]. Also according to the Euler's theorem $q^{\varphi(k)} \equiv 1 \pmod{k}$ for every $q$ coprime to $k$ that $\mathbb{Z}/k\mathbb{Z}$ has $\varphi(k)$ elements. For a prime $p$, $\varphi(p) = p - 1$ defines the number of unique motional modes. Customarily the primitive roots (mod $p$) of the $p^{th}$ cyclotomic polynomial, *i.e.*, eigenvalues are worked out by diagonalizing the system's matrix representation via unitary transformation. Circulant matrices as operators generate the cyclic group $(\mathbb{Z}/p\mathbb{Z})^\times$ where $p$ is equal to 1, 2, 4, $p^r$, $2p^r$, and $p^r$ is a power of an odd prime number [113].

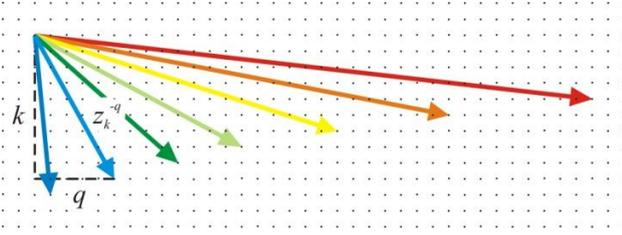

Figure 7. Energy landscape can be factorized to a discrete set of stationary states (color coded), depicted as phasors $z_k^{-q}$ where the coprime $q$ and $k$ satisfy Bézout's identity. The factorization to Euclidean domains is unique when the primitive $p^{th}$ root $z_p^{-qr} \equiv q$ (mod $p$) congruent to a power $r$ equals one on the commutative ring $\mathbb{Z}/p\mathbb{Z}$ modulo prime $p$ with unit $q$. The unique decomposition of conserved motions on closed modular orbits can be represented by the phasor sum known as the Riemann zeta function.

The analytic phasor representation is familiar from diffraction caused by an abrupt potential change. The phasors sum a pattern of alternating maxima and minima [114]. The local extrema are the stationary points where gradients vanish corresponding to the zeros of the number field. In general, the phasors track transitions from one potential to another, eventually limiting to the zero density [115]. These self-similar logarithmic spirals, e.g., galaxies, cyclones, where curvature decreases with distance so that $\ln s \partial_s \ln s$ is a constant, are ubiquitous [116,117].

The state space is represented Riemann zeta function $\zeta(s)$ of a complex variable $s = \sigma + i\tau$ and Dirichlet's character $\chi(q^{\varphi(p)}) = 1$ that may be considered, after the unique factorizing of integers $n$ to primes $p$ as a product of phasor sums

$$\zeta(s) = \sum_{n=1}^{\infty} n^{-s} = \lim_{n \to \infty}\left(\sum_{m=1}^{n} m^{-s} - \frac{n^{-s}}{1-s}\right) = \prod_{p=1}^{\infty}\left(1 - p^{-s}\right)^{-1}$$
$$= \lim_{n \to \infty}\left(\prod_{p=1}^{n}\sum_{r=0}^{p-1} p^{-(\sigma+i\tau)r} + \mathbf{O}(f(n))\right) \quad (7.1)$$
$$= \lim_{n \to \infty}\left(\prod_{p=1}^{n}\sum_{r=0}^{p-1} z_p^{-qr} + \mathbf{O}(f(n))\right) = \lim_{n \to \infty}\left(\prod_{p=1}^{n}\frac{1 - z_p^{-qp}}{1 - z_p^{-q}} + \mathbf{O}(f(n))\right)$$

which says that each term $1 - p^{-s}$ is invertible to the sum over the number field when $p^{-\sigma}$ is non-zero and not divisible by $p^{-i\tau}$. Since $|n^{-s}|$ limits to zero when $n \to \infty$, the partial sums of increasing index will limit increasingly close together as errors bounded by $f(n)$ will limit to zero [115,118,119]. Geometrically speaking the partial sums limit to a smooth curve but for $\zeta(s)$ to have zeros, it must be modular. This is in accordance with the Taniyama-Shimura conjecture that all rational elliptic curves are modular. For a given prime $p$ the partial sum $\Sigma_r z_p^{-qr} = (1 - z_p^{-qp})/(1 - z_p^{-q})$ of phasors $z_p^{-qr} = \exp(-i2\pi qr/p) = 1$ over the commutative ring limits to $\Sigma_r z_p^{-qr} = (1 - z_p^{-q})^{-1}$ and is zero at the primitive root modulo $p$ of index $r$. According to the Euclid's lemma this is the unique factorization domain of the integrally closed Noetherian ring $\mathbb{Z}/p\mathbb{Z}$ with Krull dimension one. The corresponding field $F_p(z_p^{-q})$ is a discrete valuation ring with the unique irreducible element.

The kernel basis is orthogonal $\Sigma_r z_p^{qr} z_p^{-q'r} = p\delta_{qq'}$. Thus the dot product defines, in accordance with Schur orthogonality relations, the norm $p^{-\frac{1}{2}}$ of the irreducible representation as a unitary matrix. In the same way as the discrete Fourier transformation $\Sigma_n n^{-\frac{1}{2}} z_n^{qr}$ that has the primitive $n^{th}$ root of unity $z_n = \exp(i2\pi/n)$ as its kernel, is normalized, also the number theoretic transformation $\Sigma_p p^{-\frac{1}{2}} z_p^{qr}$ mod $p$ that has $z_p = \exp(i2\pi/p)$ as its kernel, is unitary when $\sigma = \frac{1}{2}$. The unitary transformation and its inverse retain the vector lengths in accordance with Parseval's theorem. In other words the resulting map is an isometry with respect to the $L^2$ (Euclidean) norm. The result about the norm is in accordance with Plancherel theorem that energy is conserved in the discrete Fourier transform. The existence of a norm means integrally closed [stationary state]. The set of solutions for a system of polynomial equations is non-singular which means in the geometric sense that the algebraic variety is flat. The algebraic structure is determined by the C*-identity $\|\mathbf{s}\|^2 = \|\mathbf{s}^*\mathbf{s}\|$ together with the spectral radius formula. The $\|\mathbf{s}\|$ norm-imposed symmetry $\mathbf{s}^* = 1 - \mathbf{s}$ at $\zeta(\frac{1}{2}\pm i\tau) = 0$ is the conservation characteristic of partial sums [115] and is also apparent in the unitary transform $(2\pi)^{-1}\int_{-\infty}^{\infty}|\{\mathcal{M}f\}(\frac{1}{2}+i\tau)|^2 d\tau = \int_0^{\infty}|f(x)|^2 dx$ due to Mellin of a function $f$ in $L^2$.

The physical portrayal of $\zeta(\mathbf{s})$ as a number field representing the sum of quantized states means that the Riemann hypothesis $\zeta(\frac{1}{2}+i\tau) = 0$ is true [120]. The steady states are per definition conserved in energy, i.e., there are no net flows to or from the surroundings. The manifold with the $L^2$ norm is flat. The norm is necessary to diagonalize the Hermitian matrix so that the characteristic equation can be solved to yield, according to the spectral theorem, all real eigenvalues and orthogonal eigenvectors. In contrast, an evolving manifold is curved and without the norm because a change of state is inherent in net dissipation. The lack of norm, as was shown above for the Navier-Stokes equation, prevents from solving the equation, i.e., finding any zeros.

**Theorem 7.2** *All non-trivial zeros of the Riemann zeta function have real part ½.*

*Proof*. The infinite sum over integers $n$ of the Riemann zeta function $\zeta(\mathbf{s})$ can be uniquely factorized to the Euler product over all primes $p$

$$\zeta(\mathbf{s}) = \sum_{n=1}^{\infty} n^{-\mathbf{s}} = \prod_{p=1}^{\infty}\left(1 - p^{-\mathbf{s}}\right)^{-1} \quad (7.2.1)$$

as proven by Euler. The infinite sequence (Eq. 7.2.1) is denoted equivalently in the equation 7.1 as the sequence limiting to infinity because, according to the fundamental theorem of arithmetic, $n$ greater than 1 can be written as a unique product of prime numbers. Hence the partial product up to the prime $p$, when expanded, is the partial sum up to $n$ that contains those terms $n^{-\mathbf{s}}$ where $n$ is a product of primes less than or equal to $p \to \infty$. Each term in the product that follows from the Dirichlet series $\Sigma \chi_n n^{-\mathbf{s}}$ with characters $\chi_n = 1$ is invertible to the formal power series from $r = 1$ to $p - 1$ over the number field. When the product's any one term $1 - z_p^{-qp} = 0$ (Eq. 7.1) where $z_p^{-q}$ is the primitive $p^{th}$ root of unity of index $-q$, then $\zeta(\mathbf{s}) = 0$. The orthogonal kernel basis of the unique factorization

$$\sum_{r=0}^{p-1} \overline{z}_p^{-qr} z_p^{-q'r} = \sum_{r=0}^{p-1} e^{i\frac{2\pi}{p}qr} e^{-i\frac{2\pi}{p}q'r} = p\delta_{qq'} \quad (7.2.2)$$

imposes the normalization $p^{-\frac{1}{2}}$ so that the group generator, a unitary matrix $\mathbf{U}$, i.e., $\mathbf{U}^{-1} = \mathbf{U}^*$ with all eigenvalues $\lambda$ of absolute value 1, can be constructed. Hence, the condition

$$\sum_{r=1}^{p} U_{qr}^* U_{qr} = p^{-1}\sum_{r=1}^{p} p^{(\sigma+i\tau)r} p^{(\sigma-i\tau)r} = \sum_{r=1}^{p} p^{2\sigma-r} = 1 \quad (7.2.3)$$

is necessary for the zeros to exist. It defines $\sigma = \frac{1}{2}$. This choice ensures the unitarity by removing the dependence on $r$. The condition for the real part of $\mathbf{s}$ is obtained likewise from the unitary property $|\det(\mathbf{U})| = 1$, conveniently from $\det(\mathbf{U}) = \exp\{\log(\mathbf{U})\}$, or directly from $p^{-\sigma+i\tau}p^{-\sigma-i\tau} = 1/p$. It follows from the spectral theorem that $\mathbf{U}$ as unitary, can be decomposed to the canonical form of eigenvectors and eigenvalues $\lambda$ so that the characteristic equation $\det(\mathbf{I}\lambda - \mathbf{U}) = 0$ of the polynomial $\zeta(\mathbf{s}) = \Pi_p(1 - p^{-\mathbf{s}})^{-1} = 0$. Thus, it is concluded that the Riemann zeta function $\zeta(\mathbf{s})$ has all its non-trivial roots $\zeta(\frac{1}{2} + i\tau) = 0$ when the real part is ½.

The unitary condition for the zeros to exist resolves also the generalized and extended Riemann hypotheses. Physically speaking, the equation for a standing wave can be solved because for its normalized space there is a bound operator $U$ (as the representation of $\mathbf{U}$) whose resolvent $(I\lambda - U)^{-1}$ is the convergent Neumann series $\Sigma U^n$. In contrast, the motion of a non-conserved string is non-integrable, i.e., unsolvable. Furthermore, the number field presentation clarifies that the convolution of $\zeta(\mathbf{s})$ with the Möbius function is an inverse transform since there are terms in $\mu(k) = \Sigma \exp(i2\pi q/k)$, $1 \leq q \leq k$, $\gcd(q,k) = 1$ that match those in $\Sigma z_p^{-q}$, $q < p \in \mathbb{P}$ in the same way as the discrete Fourier transformation (and its inverse) picks up signal modulations to spectral lines (and vice versa) [112]. The transformation is effective in diagonalizing circulant matrices and computing cross-correlation [121]. The $L^2$ norm as the necessity for the solutions to exist sheds light on the approximations of prime-counting, e.g., by $|Li(x) - \pi(x)|$, and also on the conjecture $|M(n)| < n^{\frac{1}{2}}$ by Mertens concerning $M(n) = \Sigma \mu(k)$ that yields the Farey sequence of order $n$ which is associated to the Riemann hypothesis [122].

It is has not escaped our notice that the Riemann zeta function appears in many formula of physics – and much of its fundamental reasons have been understood [123,124, 125]. For example, since the non-trivial zeros relate to the spectral decomposition stationary states, $\zeta(\mathbf{s})$ is indeed the appropriate kernel for the integration over the stationary states. The regularization is familiar from the integration of spectral density due to Planck as well as when integrating over energy density in a tiny cavity that demonstrates Casimir effect. According to the 2$^{nd}$ law, the system is moving to abolish the Casimir force, i.e., the energy density difference between the surroundings and the parallel-plated cavity that is restricted (depleted) in its modes. Likewise, the Universe in its entirety is limiting by dilution toward the zero-density surrounding at the maximal rate $(\ln s \partial_s \ln s)^{-1}$ along least action paths when new modes of prime order appear in the expanding giant Riemannian resonator.

## 8. In between the stationary states

The representation of the energy landscape as a number field, where zeros correspond to the stationary states, encourages one to inspect the mathematical formalism also for transitions between the states. In analogy to $\zeta(s)$ (Eq. 7.1), the $L$-series $\Sigma \chi_n n^{-s}$ of Dirichlet character $\chi_n$ can be continued analytically to the $L$-function in the complex plane [126]. Likewise, $L(s)$, where $s = \sigma + i\tau$, can then be uniquely factorized to an Euler product over primes $p$

$$
\begin{aligned}
L(C,s) &= \prod_{p \in \mathbb{P}} (1 - \lambda_p)^{-1} = \lim_{n \to \infty} \left( \prod_{p \in \mathbb{P}}^{n} \sum_{q=0}^{p-1} \lambda_p^q + \mathcal{O}(f(n)) \right) \\
&= \lim_{n \to \infty} \left( \prod_{p \in \mathbb{P}}^{n} \frac{\lambda_p^p - 1}{\lambda_p - 1} + \mathcal{O}(f(n)) \right) = \lim_{n \to \infty} \left( \prod_{p \in \mathbb{P}}^{n} \Phi_p(\lambda_p) + \mathcal{O}(f(n)) \right) \\
&= \lim_{n \to \infty} \left( \prod_{p \in \mathbb{P}}^{n} \prod_{q=1}^{p-1} (\lambda_p^q - 1)(\lambda_p^p - 1) + \mathcal{O}(f(n)) \right)
\end{aligned}
\quad (8.1)
$$

so that each resolvent $(1 - \lambda_p)^{-1}$ is given as a formal power series of $\lambda_p = \chi_p p^{-s}$ of a primitive character $\chi_p$. For each series (Gaussian sum) $\Sigma \lambda_p^q$ coefficients of $(\lambda_p^p - 1)/(\lambda_p - 1) = \Phi_p \Phi_1 = (\lambda_p - 1)^{p-1}$, except the highest index $r$ are multiples of $p$ where the lowest constant equals $p$. Thus the $p^{\text{th}}$ cyclotomic polynomial $\Phi_p = \Pi(\lambda_p^q - 1)$ is found by the Eisenstein's criterion irreducible over the rational numbers [127]. At the $p^{\text{th}}$ primitive root $\lambda_p^{qr} \equiv q \pmod{p}$ congruent to a power $r$ the kernel of number theoretic transformation equals one in the ring $\mathbb{Z}/p\mathbb{Z}$ of integers modulo $p$ with unit $q$. The $p^{\text{th}}$ cyclotomic field $F_p(s)$ as a vector space splits $\Phi_p$ over rational numbers. Its Galois group is isomorphic to the multiplicative group of units of the ring $\mathbb{Z}/p\mathbb{Z}$. For a given $p$ the product's prefactor $c_p = {}_q\Pi^{p-1}(1 - \lambda_p^q) \neq 0$ because $\lambda_p^q \neq 1$ over $q = 1..p - 1$. The unitary condition of $L(s)$ is necessary for the roots to exist, just as it is for $\zeta(s)$, i.e., $\chi_p^{-2} p^{2\sigma - 1} = 1$.

The evolutionary transmission from a stationary state to another can be addressed by the Birch and Swinnerton-Dyer (BSD) conjecture that relates the order of the $L$-function's zero to the rank $r$ of an elliptic curve $C$ [128]. The number of points mod $p$ on $C$ was found to obey $\Pi_{p<x} N_p/p \sim (\log x)^r$ as $x \to \infty$. The curve $C$ is the physical medium of energy density that paves a trajectory over time, i.e., an action from one stationary state toward another.

**Theorem 8.2.** *The Taylor expansion of $L(C, s)$ at $s = 1$ has the form $L(C, s) = c(s - 1)^r +$ higher order terms with $c \neq 0$ and $r = \text{rank}(C(\mathbb{Q}))$ of the elliptic curve $C$.*

*Proof.* The $L$-series of $C$ can be factorized uniquely as in Eq. 8.1 when the appropriately weighted discriminant of $C$, $\Delta(C)$, i.e., the conductor $N$ is divisible by primes $p \parallel N$, of which $r$ denotes the highest $\gcd(r, N) = 1$. Then the modular $C$ associates with the cyclic group $(\mathbb{Z}/p\mathbb{Z})^{\times}$ [113]. At this branch $L_p(C, s) = 1 - \chi_p p^{-s}$ the Hasse-Weil zeta function $L(C, s) = \Pi L_p(C, s)^{-1}$ is amenable to the multiplicative reduction about $s = 1$ as in Eq. 8.1. The series for each $p$ that leads up to the rank $r$

$$
\begin{aligned}
\prod_{p=1}^{r} \sum_{q=0}^{p-1} \lambda_p^q &= \prod_{p \in \mathbb{P}}^{r} \frac{\lambda_p^p - 1}{\lambda_p - 1} = \prod_{p=1}^{r} \prod_{q=1}^{p-1} (\lambda_p^q - 1)(\lambda_p^p - 1) \\
&= \prod_{p=1}^{r} c_p (\lambda_p^p - 1) = c(s - 1)^r = 0
\end{aligned}
\quad (8.2)
$$

vanishes at the $p^{\text{th}}$ primitive root $\lambda_p^p = 1$. Moreover, each $\Sigma \lambda_p^q$ as irreducible $\Phi_p$ is factored to the product of the last term $(\lambda_p^p - 1)$ and the preceding terms $c_p = {}_q\Pi^{p-1}(\lambda_p^q - 1) \neq 0$ [129] because $\lambda_p^q \neq 1$ over $q = 1..p - 1$. At the roots the conjectured form $c(s - 1)^r$ is obtained when denoting $c = {}_p\Pi^r c_p$. Thus it is concluded from the unique factorization in Eq. 8.2 that the order of the zero is equal to the rank $r$ of the Abelian group of points over the field of $C$.

Conversely, the $L$-series of $C$ cannot be factorized uniquely when $p^2 \mid N$. At this branch $L_p(C, s) = 1$ the polynomial has at least one multiple root. The Euler factor of a singular point equals 1. Thus the additive reduction over nilpotent operators gives no contribution to $L(C, s = 1)$.

The $L$-series of $C$ cannot be factorized uniquely either when $p \nmid N$. At this branch $L_p(C, s) = 1 - \chi_p p^{-s} + p^{1-2s}$ where $\chi_p = 1 + p - N_p$, a residue remains in the resolvent. Since no unit is a nilpotent, it follows that there is a contributing factor to $L(C, s)$ which is given in BSD by the Taylor polynomial of the vector field coefficients about $s = 1$. This jet factor directs from the stationary state toward the other of a higher rank than $r$.

In summary, at the stationary branch where $p \parallel N$, the $L$-function as an algebraic number field has a norm and zeros exist. Accordingly, the curve with zeros (nodes) is modular in agreement with Taniyama-Shimura conjecture that all rational elliptic curves are modular. Physically speaking the conserved system is on Lyapunov-stable, closed orbits (Fig. 8). At the singular branch, where $p^2 \mid N$, squared operators, e.g., square of exterior derivate, do not contribute. Coordinates are indistinguishable due to lack of gradients in a medium without sources or sinks. A wave propagates straight at a constant frequency. At the evolutionary branch, where $p \nmid N$, there is no norm, hence there is no root either. The $L$-function of a non-modular $C$ over a finite field is non-intergrable, for the same reason as the Navier-Stokes equation has in general no solution. However Hasse's theorem sets bounds for $N_p$ by the curve that closes to a steady state at an adjacent prime. The matrix representation of the non-modular group cannot (quite) be diagonalized but Frobenius' trace departs by $|\chi_p|/2p^{½} \leq 1$ from the $L^2$ norm of modular $C$. The residual, characteristic contribution of the evolutionary branch identifies to an off-diagonal free energy element that forces the system to change its state. It is the energy flux that breaks the stationary-state symmetry of conserved currents [44]. The system evolves either by emitting from its scalar potential repository of energy or by absorbing to it.

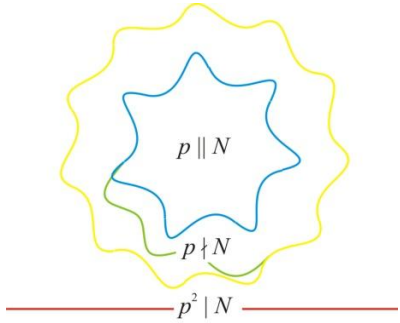

Figure 8. Portrait of the three $L$-function branches corresponding to the stationary-state periodic motions along modular trajectories $p \parallel N$ (blue and yellow), the evolutionary transition from one stationary state to another spiraling on an open track $p \nmid N$ (green) and the free propagation via alternating sense of rotation through homogenous medium $p^2 \mid N$ (red) where no coordinates can be distinguished from each other (singular).

The provided physical portrayal of the mathematical theorems is obviously no end in itself and no necessity for the pertinent proofs either but serves to materialize the significance of studies that are pictured by many merely as abstract academic exercises. The connection between the mathematical formalism and the physical law clarifies why mathematics is so powerful in rationalizing reality. However, mathematical physics has mostly specialized in computable, i.e., stationary, deterministic systems whereas in general nature's evolution is inherently intractable.

## 9. Conclusions

The present-day puzzles concerning the increasing expansion rate of the Universe as well as galactic rotation, lensing and red-shift clustering are according to the old law of thermodynamics not perplexing phenomena but follow from the conservation of energy in natural processes. At all scales, the changes in kinetic energy balance the changes in scalar and vector potentials when diverse transformations are machining the curved, universal energy density manifold toward the complete, dilute flatness of nothing. The tenet is not new [130,131] but motivated here by the universal law. The 2$^{nd}$ law, in accordance with the general theory of relativity, pictures space as the energy landscape but in addition resolves conceptual conundrums by identifying the irrevocable flow of energy from high to low densities with the irreversible flow of time.

The holistic view of nature where no system is without surroundings parallels thinking of Mach in that each subject is in relation to the common universal system. Superior surroundings impose forces by density differences in the forms spatial scalar and temporal vector potentials that dictate the evolution of any system. The natural selection for the least-time consumption of free energy does not long for some anthropic fine-tuning of process parameters [132]. The evolution of the Universe is no miraculous scenario, just as the development of an organism is neither that, but all processes keep redirecting by the mere quest to level the energy density differences in the least time. The ubiquitous principle does not recognize any demarcation line between animate and inanimate.

In this study the general principle of the maximal energy dispersal was communicated in simple terms. The deliberate lack of sophistication in addressing seemingly complicated questions may appear to some unconvincing. Admittedly the thermodynamic account on the phenomena is qualitative but within the ballpark of experimental data which itself is still subject to uncertainties. Furthermore, to substantiate the reasoning by the general principle, many more matters than were addressed or referred to deserve to be inspected by the 2$^{nd}$ law. Then the explanatory power of the universal law can be adequately contrasted with prevalent propositions, in particular with those that give a commanding role in dynamics and evolution to dark matter and dark energy. These concepts, *ad hoc* from the thermodynamic viewpoint, aim at amending the inherent impossibility to describe energy fluxes by the formalism devised for conserved systems. The defect in accounting for the open, evolving systems by mathematics of invariants has, of course, been recognized in many other contexts but perhaps progress in formulation would be now guided by the universal principle of the maximal energy dispersal given in mathematical forms. After all the objective of physics ought to be in making the world in which we live comprehensible to us, not in composing incomprehensible worlds of its own from which to live. The 2$^{nd}$ law in explaining much with little reminds us from the well-known quote of Einstein [133]. *A theory is the more impressive the greater the simplicity of its premises, the more different kinds of things it relates, and the more extended its area of applicability. Therefore the deep impression that classical thermodynamics made upon me. It is the only physical theory of universal content which I am convinced, that within the framework of applicability of its basic concepts will never be overthrown.*

**Acknowledgments**. I am grateful to Szabolcs Galambosi, Ari Lehto, Dharmindar Maharaj, Heikki Sipilä, Donald Spector and Tuomo Suntola for insightful conversations.

## Appendix

*The probability measure of a system* (Eq. 2.2) is derived by considering transitions between states indexed with *j* and *k* as pictured in Fig. A1. Energy embodies the states. When the *k*-state absorbs quanta from its surroundings (or emits quanta to its surroundings), the *j*-state emerges. Thus the probability $P_j$ of the *j*-entity (Eq. 2.1) depends on the number $N_k$ of *k*-entities and the energy difference $\Delta G_{jk} = G_j - G_k$ between the *j* and *k* states and the dissipation $\Delta Q_{jk}$ to (or from) the surroundings relative to the average energy density $k_BT$. These ingredients of the *jk*-transition are considered as statistically independent hence given as a product (Eq. 2.1). The degeneracy $g_{jk}$ numbers the *k*-entities that are indistinguishable (symmetric) within each indistinguishable *j*-entity. Finally, each entity is regarded as a system of its own. Systems within system form the total system characterized by the additive logarithmic probability measure $\ln P = \Sigma \ln P_j$.

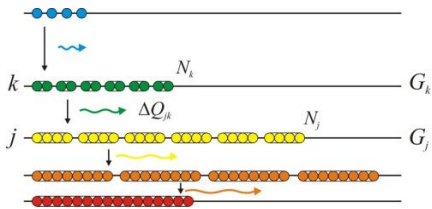

Figure A1. Energy level diagram of a system houses a diversity of entities. A pool of *j*-entities (colored aggregates), in numbers $N_j$, is associated with an energy density $\phi_j = N_j\exp(G_j/k_BT)$. According to the 2$^{nd}$ law the system evolves from a state to another, more probable one by diverse dissipative (color-coded waves) *jk*-transformations that diminish the energy density differences, i.e., the free energy $\Sigma k_BT\ln\phi_k + \Delta Q_{jk} - k_BT\ln\phi_j$ until the stationary state in the respective surroundings has been attained.

*The equation of evolution in the continuum* (Eq. 3.1) is derived from Eq. 2.2 by considering transitions between continuum states indexed with *j* and *k*. The steady-state equation $N_jk_BT = p_jV = \int \mathbf{F}\cdot d\mathbf{x}$ for the (ideal gas) *j*-system equates the kinetic energy $2K$ with the potential *U*. Particularly in a stationary system $\partial_t\ln P_j = 0$, $\ln P_j = N_j$, the entities $N_j$ generate a pressure $p_j = F_j/A$ due to mutual forces $F_j$ = $\Sigma \partial_t(m_{jk}v_k) = \Sigma v_j\partial_{xj}(m_{jk}v_k)$ per area $A = \int d^2x$ in a volume $V = \int d^3x$ to yield $\Sigma x_jF_j = \Sigma x_jm_{jk}a_k = -U$ in accordance with the virial theorem of a conserved system ($dm_{jk} = 0$). Generally in an evolving system $\partial_t\ln P_j > 0$

$$2K = \sum_{j,k} x_j d_t\left(m_{jk}v_k\right) = \sum_{j,k} x_j m_{jk}a_k + \sum_{j,k} x_j \partial_t m_{jk} v_k \quad\quad (A.1)$$
$$= \sum_{j,k} x_j m_{jk}a_k + \sum_{j,k} \int v_j \partial_t E_{jk} v_k / c^2 \, dt = -U + Q$$

where changes in kinetic $d_t2K = d_t(\Sigma v_k m_{jk}v_k) \equiv k_BTd_t\ln P$, potential $\partial_tU = -\Sigma v_j m_{jk}a_k$ and dissipation $\partial_tQ = \Sigma v_j\partial_t m_{jk}v_k = \Sigma v_j\partial_t E_{jk}v_k/c^2 = \partial_tE/n^2$ have been integrated.

When the 4-gradient $\partial_\mu = (\partial_t/c, \partial_x, \partial_y, \partial_z)$ acts on the 4-potential $A_\mu = (-U, Q_x, Q_y, Q_z)$, continuity is expressed by the field equation

$$F_{\mu\nu} = \partial_\mu A_\nu - \partial_\nu A_\mu = \begin{pmatrix} 0 & -F_x & -F_y & -F_z \\ F_x & 0 & R_z & -R_y \\ F_y & -R_z & 0 & R_x \\ F_z & R_y & -R_x & 0 \end{pmatrix} \quad (A.2)$$

where $\partial_t\mathbf{p} = \mathbf{F} = -\nabla U + \partial_t\mathbf{Q}/c$ and $\mathbf{R} = \nabla \times \mathbf{Q}$. The open, evolving system as the manifold in motion, is represented by the tensor

$$d_t 2K_{\mu\nu} = v_{\mu\nu}F^{\mu\nu} = \begin{pmatrix} 0 & -v_x\partial_xU & -v_y\partial_yU & -v_z\partial_zU \\ \partial_tQ_x & 0 & -v_yR_z & v_zR_y \\ \partial_tQ_y & v_xR_z & 0 & -v_zR_x \\ \partial_tQ_z & -v_xR_y & v_yR_x & 0 \end{pmatrix} \quad (A.3)$$

where the 4-vector velocity $v_\mu = (-c, v_x, v_y, v_z)$. The (power) tensor for the energy flows contracts to the 0-form $d_t2K = \Sigma d_t2K_{\mu\nu} = -\mathbf{v}\cdot\nabla U + \partial_t\mathbf{Q} + \mathbf{v}\times\mathbf{R}$. The change in kinetic energy reports from the changes in scalar potential as well as from the changes in vector potential due flows of light and matter to the open system from the surroundings or vice versa.

*The inverse distance dependence of the scalar potential* follows from the stationary action of a conserved system: $2Kt + Ut$ is a constant. The stationary path (closed orbit) means $\partial_t(2Kt) + \partial_t(Ut) = 0 \Leftrightarrow 2K + U = -t\partial_t2K + t\partial_tU$ at all times. Since $dm = 0$ the condition $2K + U = 0$ yields $mv^2 = -U$ and $\partial_t2K + \partial_tU = 0$ yields $ma = -\partial_rU$ using $dr = vdt$. The combined condition for the centro-symmetric potential $r\partial_rU = -U$ is satisfied by $U(r) \propto 1/r$. The proportionality coefficient *G* can be determined, e.g., from the divergence of the field $\nabla\cdot\mathbf{r}/r^3 = G\rho$, whose source is, e.g., a density $\rho = M\delta(r)$ of a point-like source *M* at $r = 0$.

*The frequency shift of radiation in refraction* follows from the conservation of energy when light crosses from an energy density *k* to another *j* (Fig. A2). Energy is conserved

$$d_t2K = -\partial_tU + \partial_tQ \Leftrightarrow d_tmv^2 = -\partial_t\frac{GmM_k}{r} + \partial_tmc^2 \quad (A.4)$$
$$\Rightarrow \frac{1}{n^2} = \frac{v^2}{c^2} = 1 - \frac{GM_k}{c^2r} \Rightarrow d_t2K_k = d_t2K_j \Rightarrow f_k = \frac{n_j}{n_k}f_j$$

where the index of refraction $n_k^2 = c^2/v_k^2 = (1 - GM_k/c^2 r_k)^{-1}$ contains the ratio of the universal $R = c^2/a_t$ to local $r_k = v_k^2/a_k$ radii of curvatures due to the energy density embodied in the respective universal $M = \Sigma m_i$ and local $M_k$ sources. Consistently with the principle of least action á la Fermat, i.e., the law due to Snell, the ray of light traces past the $k$-locality along the path where its frequency shifts according to Eq. A.2 due the changing index of refraction.

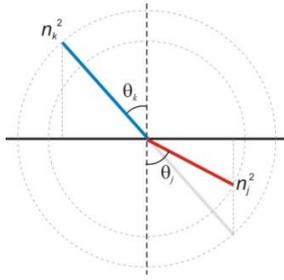

Figure A2. Refraction of light at an energy density boundary specified by indexes $n_k^2 > n_j^2$ that relate to curvatures $1/r_k$ and $1/r_j$ (dashed circles) relative to the universal curvature $1/R$ of reference index $n^2 = c^2 \varepsilon_o \mu_o = 1$. The color-coded frequencies $f_k$ (high) and $f_j$ (low) relate to each other by Eq. A2. For a certain energy $hf_k$ there is no partition beyond the limit of total internal reflection $\theta_j = \pi/2$ at $n_k^2 \sin^2\theta_k > n_j^2$ because the energy density in the $j$-medium as a resonator falls short to support the propagation. Thus the lowest frequencies deflect least and distribute most uniformly.

## References

1. Goldhaber, G. & Perlmutter, S. 1998 A study of 42 type Ia supernovae and a resulting measurement of Omega(M) and Omega(Lambda). *Physics Reports-Review Section of Physics Letters* **307**, 325–331.
2. Riess, A. G., Filippenko, A. V., Challis, P. & et al. 1998 Observational evidence from supernovae for an accelerating universe and a cosmological constant. *Astron. J.*, **116**, 1009–1038.
3. Garnavich, P. M., Kirshner, R. P., Challis, P. & et al. 1998 Constraints on cosmological models from Hubble Space Telescope observations of high-z supernovae. *Astrophys. J.* **493**, L53+ Part 2.
4. Garnavich, P. M., Kirshner, R. P., Challis, P. & et al. 1998 Constraints on cosmological models from Hubble Space Telescope observations of high-z supernovae. *Astrophys. J.* **493**, L53+ Part 2.
5. Volders, L. 1959 Neutral hydrogen in M 33 and M 101. *Bulletin of the Astronomical Institutes of the Netherlands* **14**, 323–334.
6. De Blok, W. J. G. & McGaugh, S. 1997 The dark and visible matter content of low surface brightness disc galaxies. *MNRAS* **290**, 533–552.
7. Clowe, D., Bradač, M., Gonzalez, A. H., Markevitch, M., Randall, S. W., Jones, C. & Zaritsky, D. 2006 A direct empirical proof of the existence of dark matter. *Astrophys. J.* **648**, L109–L113. (doi: 10.1086/508162)
8. Webb, J. K., Flambaum, V. V., Churchill, C. W., Drinkwater, M. J. & Barrow, J. D. 1999 Search for time variation of the fine structure constant. *Phys. Rev. Lett.* **82**, 884–887. (doi:10.1103/PhysRevLett.82.884) (arXiv:astro-ph/9803165)
9. Webb, J. K., Murphy, M. T., Flambaum, V. V., Dzuba, V. A., Barrow, J. D., Churchill, C. W., Prochaska, J. X. & Wolfe, A. M. 2001 Further evidence for cosmological evolution of the fine structure constant. *Phys. Rev. Lett.* **87**, 091301. (doi:10.1103/PhysRevLett.87.091301)
10. Khatri, R. & Wandelt, B. D. 2007 21-cm radiation: A new probe of variation in the fine-structure constant. *Phys. Rev. Lett.* **98**, 111301. (doi:10.1103/PhysRevLett.98.111301)
11. Paal, G. 1970 The global structure of the universe and the distribution of quasi-stellar objects. *Acta Phys. Acad. Sci. Hung.* **30**, 51–54.
12. Tifft, W. G. 1973 Properties of the redshift-magnitude bands in the Coma cluster. *Astrophys. J.* **179**, 29–44.
13. Tifft, W. G. 2003 Redshift periodicities, the galaxy-quasar connection. *Astrophys. Space Sci.* **285**, 429–449.
14. Carnot, S. 1824 Reflexions sur la puissance motrice du feu et sur les machines propres a developper cette puissance. Paris, France: Bachelier.
15. Jaynes, E. T. 2003 *Probability theory. The logic of science.* Cambridge, UK: Cambridge University Press.
16. Sharma, V. & Annila, A. 2007 Natural process – Natural selection. *Biophys. Chem.* **127**, 123–128. (doi:10.1016/j.bpc.2007.01.005)
17. Du Châtelet, E. 1740 *Institutions de physique*. Paris, France: Prault. Facsimile of 1759 edition *Principies mathématiques de la philosophie naturelle I-II*. Paris, France: Éditions Jacques Gabay.
18. 's Gravesande, W. 1720 *Physices elementa mathematica, experimentis confirmata, sive introductio ad philosophiam Newtonianam*. Leiden, The Netherlands.
19. Jaakkola, S., Sharma, V. & Annila, A. 2008 Cause of chirality consensus. *Curr. Chem. Biol.* **2**, 53–58. (doi:10.2174/187231308784220536) (arXiv:0906.0254)
20. Jaakkola, S., El-Showk, S. & Annila, A. 2008 The driving force behind genomic diversity. (arXiv:0807.0892)
21. Grönholm, T. & Annila, A. 2007 Natural distribution. *Math. Biosci.* **210**, 659–667. (doi:10.1016/j.mbs.2007.07.004)
22. Würtz, P. & Annila, A. 2008 Roots of diversity relations. *J. Biophys.* (doi:10.1155/2008/654672) (arXiv:0906.0251)
23. Annila, A. & Annila, E. 2008 Why did life emerge? *Int. J. Astrobiol.* **7**, 293–300. (doi:10.1017/S1473550408004308)
24. Annila, A. & Kuismanen, E. 2008 Natural hierarchy emerges from energy dispersal. *BioSystems* **95**, 227–233. (doi:10.1016/j.biosystems.2008.10.008)
25. Karnani, M. & Annila, A. 2009 Gaia again. *BioSystems* **95**,

82–87. (doi: 10.1016/j.biosystems.2008.07.003)
26. Sharma, V., Kaila, V. R. I. & Annila, A. 2009 Protein folding as an evolutionary process. *Physica A* **388**, 851–862. (doi:10.1016/j.physa.2008.12.004)
27. Würtz, P. & Annila, A. 2010 Ecological succession as an energy dispersal process. *BioSystems* **100**, 70–78.
28. Annila, A. & Salthe, S. 2009 Economies evolve by energy dispersal. *Entropy* **11**, 606–633.
29. Annila, A. 2011 Least-time paths of light. *MNRAS* **416**, 2944–2948.
30. Koskela, M. & Annila, A. 2011 Least-time perihelion precession. *MNRAS* **417**, 1742–1746. arXiv:1009.1571
31. Tuisku, P., Pernu, T. K. & Annila, A. 2009 In the light of time. *Proc. R. Soc. A.* **465**, 1173–1198. (doi:10.1098/rspa.2008.0494)
32. Annila, A. 2010 The 2$^{nd}$ law of thermodynamics delineates dispersal of energy. *Int. Rev. Phys.* **4**, 29–34.
33. Kaila, V. R. I. & Annila, A. 2008 Natural selection for least action. *Proc. R. Soc. A.* **464**, 3055–3070. (doi:10.1098/rspa.2008.0178)
34. Annila, A. 2010 All in action. *Entropy* **12**, 2333–2358.
35. Eddington, A. S. 1928 *The nature of physical world*. New York, NY: MacMillan.
36. Gibbs, J. W. 1993–1994 *The scientific papers of J. Willard Gibbs*. Woodbridge, CT: Ox Bow Press.
37. Wyld, H. W. 1961 Formulation of the theory of turbulence in an incompressible fluid. *Ann. Phys.* **14**, 143–165.
38. Keldysh, L. V. 1964 *Zh. Eksp. Teor. Fiz.* **47**, 1515–1527; 1965 Diagram technique for nonequilibrium processes. *Sov. Phys. JETP* **20**, 1018–1026.
39. Boltzmann, L. 1905 *Populäre Schriften*. Leipzig, Germany: Barth. [Partially transl. Theoretical physics and philosophical problems by B. McGuinness, Reidel, Dordrecht 1974.]
40. De Donder, Th. 1936 *Thermodynamic theory of affinity: A book of principles*. Oxford, UK: Oxford University Press.
41. Kullback, S. 1959 *Information theory and statistics*. New York, NY: Wiley.
42. Salthe, S. N. 2002 Summary of the principles of hierarchy theory. *General Systems Bulletin* **31**, 13–17.
43. Salthe, S. N. 2007 The natural philosophy of work. *Entropy* **9**, 83–99.
44. Noether, E. 1918 Invariante Variationprobleme. *Nach. v.d. Ges. d. Wiss zu Goettingen, Mathphys. Klasse* 235–257; English translation Tavel, M. A. 1971 Invariant variation problem. *Transp. Theory Stat. Phys.* **1**, 183–207.
45. Jaynes, E. T. 1957 Information theory and statistical mechanics. *Phys. Rev.* **106**, 620–630.
46. Roshdi, R. 1992 *Optique et Mathematiques: Recherches sur L'Histoire de la Pensee Scientifique en Arabe*. Aldershot, UK: Variorum.
47. Lotka, A. J. 1922 Natural selection as a physical principle. *Proc. Natl. Acad. Sci.* **8**, 151–154.
48. Zinn-Justin, J. 2002 *Quantum field theory and critical phenomena*. New York, NY: Oxford University Press.
49. Peskin, M. & Schroeder, D. 1995 *An introduction to quantum field theory*. Philadelphia, PA: Westview Press.
50. Thirring, W. 1970 Systems with negative specific heat. *Z. Physik* **235**, 339–352.
51. Pokorski, S. 1987 *Gauge field theories*. Cambridge, UK: Cambridge University Press.
52. Darwin, C. 1859 *On the origin of species*. London, UK: John Murray.
53. Penzias, A. A. & Wilson R. W. 1965 A measurement of excess antenna temperature at 4080 Mc/s. *Astrophys. J.* **142**, 419–421. (doi:10.1086/148307)
54. Karnani, M., Pääkkönen, K. & Annila, A. 2009 The physical character of information. *Proc. R. Soc. A.* **465**, 2155–2175. (doi:10.1098/rspa.2009.0063)
55. Poynting, J. H. 1920 *Collected scientific papers*. London, UK: Cambridge University Press.
56. Lavenda, B. H. 1985 *Nonequilibrium statistical thermodynamics*. New York, NY: John Wiley & Sons.
57. Gouy, L. G. 1889 Sur l'energie utilizable. *J. de Physique* **8**, 501–518.
58. Stodola, A. 1910 *Steam and gas turbines*. New York, NY: McGraw-Hill.
59. Weinberg, S. 1972 *Gravitation and cosmology, principles and applications of the general theory of relativity*. New York, NY: Wiley.
60. Berry, M. 2001 *Principles of cosmology and gravitation*. Cambridge, UK: Cambridge University Press.
61. Taylor, E. F. & Wheeler, J. A. 1992 *Spacetime physics*. New York, NY: Freeman.
62. Lorenz, L. 1867 On the identity of the vibrations of light with electrical currents. *Philos. Mag.* **34**, 287–301.
63. Einstein, A. 1905 On the electrodynamics of moving bodies. *Annalen der Physik* **17**, 891–921.
64. Connes, A. 1994 *Noncommutative geometry (Géométrie non commutative)*. San Diego, CA: Academic Press.
65. Lemaître, G. 1927 Un univers homogène de masse constante et de rayon croissant rendant compte de la vitesse radiale des nébuleuses extra-galactiques. *Annales de la Société Scientifique de Bruxelles* **47**, 49–56.
66. Hubble, E. 1929 A relation between distance and radial velocity among extra-galactic nebulae. *Proc. Natl. Acad. Sci. USA* **15**, 168–173.
67. Hawking, S. W. 1974 Black hole explosions. *Nature* **248**, 30–31.
68. Sciama, D. W. 1953 On the origin of inertia. *MNRAS* **113**, 34–42.
69. Dicke, R. H. 1961 Dirac's cosmology and Mach's principle. *Nature* **192**, 440–441. (doi:10.1038/192440a0)

70. Haas, A. 1936 An attempt to a purely theoretical derivation of the mass of the universe. *Phys. Rev.* **49**, 411–412.
71. Tryon, E. P. 1973 Is the universe a vacuum fluctuation? *Nature* **246**, 396–397.
72. Planck, M. 1899 Über irreversible Strahlungsvorgänge. *Sitzungsberichte der Königlich Preußischen Akademie der Wissenschaften zu Berlin* **5**, 440–480.
73. Bennett, C.L. et al., 2003 First-year Wilkinson Microwave Anisotropy Probe (WMAP) Observations: Preliminary maps and basic results. *Astrophys. J. Suppl. Series*, **148**, 1–27.
74. Salthe, S. N. 1985 *Evolving hierarchical systems: Their structure and representation*. New York, NY: Columbia University Press.
75. Tully, R. B. & Fisher, J. R. 1977 A new method of determining distances to galaxies. *Astronomy and Astrophysics* **54**, 661–673.
76. Milgrom, M. 1983 A modification of the Newtonian dynamics as a possible alternative to the hidden mass hypothesis. *Astrophys. J.* **270**, 365–370.
77. Anderson, J. D., Laing, P. A., Lau, E. L., Liu, A. S., Nieto, M. M. & Turyshev, S. G. 1998 Indication, from Pioneer 10/11, Galileo, and Ulysses data, of an apparent anomalous, weak, long-range acceleration. *Phys. Rev. Lett.* **81**, 2858–2861. (doi:10.1103/PhysRevLett.81.2858)
78. Bertolami, O. Páramos J. 2004 Pioneer's Final Riddle. arXiv:gr-qc/0411020.
79. Anderson, J. D. & Nieto, M. M. 2009 Astrometric solar-system anomalies. In *Relativity in Fundamental Astronomy*. Proceedings IAU Symposium No. 261, Klioner, S. A., Seidelman, P. K. & Soffel, M. H. eds. arXiv:0907.2469v2.
80. Francisco, F., Bertolami, O., Gil, P. J. S., Páramos J. 2011 Modelling the reflective thermal contribution to the acceleration of the Pioneer spacecraft. arXiv:1103.5222.
81. Koschmieder, E. L. 1993 *Bénard cells and Taylor vortices* New York, NY: Cambridge University Press.
82. Taylor, G. I. 1923 Stability of a viscous liquid contained between two rotating cylinders. *Phil. Trans. R. Soc. A* **223**, 289–343. (doi:10.1098/rsta.1923.0008)
83. Feynman, R. P. 1955 Application of quantum mechanics to liquid helium. *Prog. Low Temp. Phys.* **1**, 17–53. (doi:10.1016/S0079-6417(08)60077-3)
84. De Maupertuis, P. L. M. 1744 Accord de différentes loix de la nature qui avoient jusqu'ici paru incompatibles. *Mém. Ac. Sc. Paris* 417–426.
85. Clemence, G. M. 1947 The relativity effect in planetary motions. *Rev. Mod. Phys.* **19**, 361–364. (doi:10.1103/RevModPhys.19.361)
86. Zwicky, F. 1933 Die Rotverschiebung von extragalaktischen Nebeln. *Helvetica Physica Acta* **6**, 110–127.
87. Zwicky, F. 1937 On the masses of nebulae and of clusters of nebulae. *Astrophys. J.* **86**, 217–246. (doi:10.1086/143864)
88. Charlton, J. C. & Schramm, D. N. 1986 Percolation of explosive galaxy formation. *Astrophys. J.* Part 1 **310**, 26.
89. Bok, B. J. & Reilly, E. F. 1947 Small dark nebulae. *Astrophys. J.* **105**, 255–257.
90. Annila, A. 2011 Physical portrayal of computational complexity. *ISRN Computational Mathematics* 321372, 1–15. (doi:10.5402/2012/321372) (arXiv:0906.1084).
91. Poincaré, J. H. 1890 Sur le problème des trois corps et les équations de la dynamique. Divergence des séries de M. Lindstedt. *Acta Mathematica* **13**, 1–270.
92. Strogatz, S. H. 2000 *Nonlinear dynamics and chaos with applications to physics, biology, chemistry and engineering*. Cambridge, MA: Westview.
93. Newman, J. R. 1956 *The world of mathematics*. New York, NY: Simon and Schuster.
94. www.claymath.org/millennium/Navier-Stokes_Equations
95. Van Moerbeke, P. 1989 Introduction to algebraic integrable systems and their Panlevé analysis. *Proceeding of Symposia in Pure Mathematics* 49.
96. Pernu, T. K. & Annila, A. 2012 Natural emergence. *Complexity* **17**, 44–47. (doi:10.1002/cplx.21388).
97. Mackay, A. L. 1977 *The harvest of a quiet eye - a selection of scientific quotations*. Bristol, UK: The Institute of Physics.
98. Kondepudi, D. & Prigogine, I. 1998 *Modern thermodynamics*. New York, NY: Wiley.
99. Verhulst, P. F. 1845 Recherches mathématiques sur la loi d'accroissement de la population. *Nouv. Mém. Acad. Roy. Sci. Belleslett. Bruxelles* **18**, 1–38.
100. May, R. M. 1976 Simple mathematical models with very complicated dynamics. *Nature* **261**, 459–467.
101. Waage, P. & Guldberg, C. M. 1864 Forhandlinger 35 Videnskabs-Selskabet i Christiana.
102. Cooley, J. W. & Tukey, J. W. 1965 An algorithm for the machine calculation of complex Fourier series. *Math. Comput.* **19**, 297–301. (doi:10.2307/2003354)
103. Lehto, A. 2009 On the Planck scale and properties of matter. *Nonlinear Dyn.* **55**, 279–298. doi:10.1007/s11071-008-9357-z
104. Alonso, M. & Finn, E. J. 1983 *Fundamental university physics*. 3. Reading, MA: Addison-Wesley.
105. Ryden, B. 2003 *Introduction to cosmology*. New York, NY: Addison-Wesley.
106. Crawford, F. S. 1968 *Waves* in *Berkeley physics course*. 3. New York, NY: McGraw-Hill.
107. Tifft, W. G. 1996 Global redshift periodicities and periodicity structure. *Astrophys. J.* **468**, 491–518.
108. Napier, W. M. 2003 A statistical evaluation of anomalous redshift claims. *Astrophys. Space Sci.* **285**, 419–427.
109. Chwolson, O. 1924 Über eine mögliche Form fiktiver Doppelsterne. *Astronomische Nachrichten* **221**, 329–330. (doi:10.1002/asna.19242212003)

110. Einstein, A. 1936 Lens-like action of a star by the deviation of light in the gravitational field. *Science* **84**, 506–507. (doi:10.1126/science.84.2188.506)
111. Rosen, K. H. 1993 *Elementary number theory and its applications*. Reading, MA: Addison-Wesley.
112. Schramm, W. 2008 The Fourier transform of functions of the greatest common divisor. *Integers* **8**, A50.
113. Richman, F. 1971 *Number theory: An introduction to algebra*. Belmont, CA: Wadsworth.
114. Feynman, R., Leighton, R. B. & Sands, M. 1963 *The Feynman lectures on physics*. Reading, MA: Addison-Wesley.
115. Erickson, C. 2005 A geometric perspective on the Riemann zeta function's partial sums. *SURJ Mathematics* 22–31.
116. Bernoulli, J. 1967 *Opera, Tomus Secundus*. Brussels, Belgium: Culture et Civilisation.
117. Mäkelä, T. & Annila, A. 2010 Natural patterns of energy dispersal. *Phys. Life. Rev.* **7**, 477–498.
118. Hardy, G. H. & Littlewood, J. E. 1921 The zeros of Riemann's zeta function on the critical line. *Math. Z.* **10**, 283–317.
119. Titchmarsh, E. C. 1986 The theory of the Riemann zeta function. Oxford, UK: Oxford Science Publications.
120. www.claymath.org/millennium/Riemann_Hypothesis/
121. Golub, G. H. & Van Loan, C. F. 1996 *Matrix computations*. Baltimore, MD: The Johns Hopkins University Press.
122. Franel, J. & Landau, E. 1924 Les suites de Farey et le problème des nombres premiers. *Göttinger Nachr.* 198–206.
123. Montgomery, H. L. 1973 The pair correlation of zeros of the zeta function. *Analytic number theory*, *Proc. Sympos. Pure Math.*, XXIV, 181–193. Providence, RI: Am. Math. Soc.
124. Berry, M. V. & Keating, J. P. 1999 H = xp and the Riemann zeros. *Supersymmetry and trace formulae: Chaos and disorder.* 355–367 eds. Keating, J. P., Khmelnitski, D. E., Lerner, I. V. New York, NY: Plenum.
125. Spector, D. 1990 Supersymmetry and the Möbius inversion function. *Commun. Math. Phys.* **127**, 239–252.
126. Silverman, J. H. 1992 *The arithmetic of elliptic curves*. New York, NY: Springer-Verlag.
127. Lang, S. 1992 *Algebra*. Reading, MA: Addison-Wesley.
128. www.claymath.org/millennium/Birch_and_Swinnerton-Dyer_Conjecture
129. Tate, J. 1974 The arithmetic of elliptic curves. *Invent. Math.* **23**, 179–206.
130. Eddington, A. S. 1931 Preliminary note on the masses of the electron, the proton and the Universe. *Proc. Camb. Phil. Soc.* **27**, 15–19.
131. Dirac, P. A. M. 1938 A new basis for cosmology. *Proc. R. Soc. A* **165**, 199–208. (doi:10.1098/rspa.1938.0053)
132. Barrow, J. D. & Tipler, F. J. 1986 *The anthropic cosmological principle*. New York, NY: Oxford University Press.
133. Klein, M. J. 1967 Thermodynamics in Einstein's thought. *Science* **157**, 509–516.